\documentclass[journal,twoside]{IEEEtran}
\usepackage{graphicx}
\usepackage{epstopdf}
\usepackage{placeins}
\usepackage{array}
\DeclareGraphicsExtensions{.eps,.pdf} 
\graphicspath{ {figures/} }
\usepackage{cite}
\usepackage{balance}
\usepackage{amssymb,amsmath}
\usepackage{subfigure}
\usepackage{amssymb,amsmath}
\usepackage{algorithm}
\usepackage{algorithmic}
\usepackage{color}
\usepackage{multirow}
\makeatletter
\newcommand\restartchapters{\par
  \setcounter{chapter}{0}%
  \setcounter{section}{0}%
  \gdef\@chapapp{\chaptername}%
  \gdef\thechapter{\@arabic\c@chapter}}
\makeatother

\newtheorem{remark}{\bf Remark}

\newtheorem{proposition}{\bf Proposition}

\newcommand{\tr}{{\mathrm{tr}}}

\newcommand{\st}{{\mathrm{subject\ to}}}

\newcommand{\lbn}{\par\medskip\noindent}

\newcommand{\ds}{\displaystyle}

\begin{document}
\bstctlcite{IEEEexample:BSTcontrol}

\title{ Spectral Efficiency of Full-Duplex Multiuser System: Beamforming Design, User Grouping, and Time Allocation }
\author{
\IEEEauthorblockN{Van-Dinh Nguyen, \textit{Student Member, IEEE,} Hieu V. Nguyen, Chuyen T. Nguyen\\ and Oh-Soon Shin, \textit{Member, IEEE}}
\thanks{V.-D.~Nguyen, H.~V.~Nguyen, and O.-S.~Shin are with the School of Electronic Engineering $\&$ Department of  ICMC Convergence Technology, Soongsil University, Seoul 06978, Korea  (e-mail: \{nguyenvandinh, hieuvnguyen, osshin\}@ssu.ac.kr).}
\thanks{C. T. Nguyen is with School of Electronics and Telecommunications, Hanoi University of Science and Technology, 1 Dai Co Viet, Hanoi, Vietnam (e-mail: chuyen.nguyenthanh@hust.edu.vn).}
}

\maketitle
%\thispagestyle{empty}
%\pagestyle{empty}
%\vspace*{-2cm}
\begin{abstract}
Full-duplex (FD) systems have emerged as an essential enabling technology  to further increase the data rate of wireless
communication systems. The key idea of FD is to serve multiple users over the same bandwidth with a base station (BS) that can
simultaneously transmit and receive the signals. The most challenging issue in designing an FD system is  to address both the harmful effects
of residual self-interference  caused by the transmit-to-receive antennas at the BS as well as the co-channel interference 
from an uplink user (ULU) to a downlink user (DLU). An efficient solution to these problems  is to assign
the ULUs/DLUs in different groups/slots, with each user served in multiple groups. Hence, this paper studies the joint design of transmit beamformers,  ULUs/DLUs group assignment, and time allocation for each group. The specific aim is to maximize the
sum rate under the ULU/DLU minimum throughput constraints. The utility function of interest is a difficult nonconcave problem, and the involved
constraints are also nonconvex, and so this is a computationally troublesome problem. To solve this optimization problem, we
propose a new path-following algorithm for computational solutions to  arrive at  least the local optima. Each iteration
involves only a simple convex quadratic program. We prove that the proposed algorithm iteratively improves the objective
while guaranteeing convergence. Simulation results confirm the fast convergence of the proposed algorithm with substantial
performance improvements over existing approaches.
\end{abstract}

\section{Introduction} \label{Introduction}

Over the last decade, significant efforts have been expended on   improving the spectral efficiency of wireless communication
systems to meet heterogeneous networking demands, including a high data rate, high reliability, and massive connectivity. Among these efforts, multi-antenna communications  have  been proposed as an essential enabling technique for many wireless communication standards \cite{MietznerCST09,Muirhead16}.
Due to the practical limitations of hardware designs, a base station (BS) is currently designed to operate in the half-duplex (HD) mode, i.e., the BS can only transmit or receive over a specific  frequency band.
However, conventional HD can no longer provide substantial improvements for  given finite radio resources. On the other hand, full-duplex (FD) is
designed to transmit information intended for downlink users (DLUs)  and to receive  information  from  uplink users (ULUs)  on the same frequency band. Thus, an FD system has been shown to greatly improve the system throughput compared to its HD counterpart \cite{Saetal14,ZhangCM15}. 

A major barrier for FD radio is the significant effects of residual self-interference (SI) caused by the transmit
antennas to the receive antennas at the BS. To be specific, the  signals transmitted for the downlink channel corrupt the desired signals in the uplink channel since both transmit and receive antennas are co-located and function at the same time and on the same frequency band.
In recent years, advances in hardware design have allowed the SI to be effectively suppressed at a reasonable cost and  it has been shown that  FD radio
may be deployable in next-generation networks  \cite{DUPLO,Duarte:TWC:12}. A wide range of residual SI mitigation  techniques have been reported in \cite{Everett:14:TWC,Saetal14} and \cite{Riihonen-SP-11}. In addition, the spectral efficiency of the FD system is also degraded by co-channel interference (CCI) caused to a DLU by the transmit signal of a ULU.

\subsection{Related Works}
 Several efforts have been carried out in various  multiuser multi-input multi-output (MU-MIMO) contexts in  FD systems to improve the overall spectral efficiency of the downlink  and uplink  channels.  Among such,  Nguyen \textit{et al.} proposed a low-complexity precoding design without CCI in \cite{Dan-SP-13} and then extended to the case of existing CCI in \cite{Dan:TWC:14}. In response,    Tam \textit{et al.} \cite{Tam:TCOM:16} studied FD communication systems in MIMO multicell networks and proposed an iterative, low-complexity successive convex quadratic programming (SCQP) algorithm to solve the optimization problem of the sum rate maximization (SRM).  To further improve the spectral efficiency, the FD was incorporated into massive MIMO in small cell wireless systems \cite{LiTVT16}.

More recently, the user selection and channel assignment have been proposed to mitigate the  effects of residual SI and CCI. Specifically,  ULUs and DLUs are selected to form two groups of users with respect to the  effects of residual SI and CCI as much as the sum rate (SR) increases \cite{NguyenICC16}.  Each user is allowed access to two orthogonal channels, one for its uplink transmission and the other for its downlink reception. Next,  the problem of grouping users into pairs and assigning different frequency channels to each pair to improve the spectral efficiency has been considered in \cite{
SilvaTWC16}. However, these approaches are  demanding in terms of the infinite radio resources due to excessive  frequency channel requirements. Notably,  Ahn \textit{et al.} \cite{AhnTWC16} proposed a low-complexity user selection method to select the best users for a given  number of users, which in turn improves the total sum spectral efficiency. The number of served users in \cite{AhnTWC16} should be noted to be very limited when the number of users becomes large.

\subsection{ Motivation and Contributions}
In this paper, we study the potential of user grouping in FD systems  to further improve the spectral  efficiency. The BS is equipped with multiple antennas and operates in the FD mode, while each user is equipped with a single antenna and  operates in the HD mode.  In contrast with the previous works in \cite{NguyenICC16,Dan:TWC:14} and \cite{SilvaTWC16},  communication is carried out in multiple time slots and over the same frequency band, as inspired by the work in \cite{RazaviyaynTSP14}. Intuitively, when the residual SI and CCI become large, the BS and ULUs need to scale down their transmit
power to satisfy  the quality-of-service (QoS) constraints, resulting in a loss in  system performance. By this very nature, we aim to assign users into multiple groups, and each group is served in  one separate time slot, which differs from a traditional grouping method since each user can only be served in one group. This helps mitigate the harmful effects of residual SI and CCI and allows us to exploit  multiuser diversity gain in both directions.  We are concerned with the problem of jointly designing  downlink beamformers, uplink transmit power allocation, user grouping, and time allocation  to maximize the spectral efficiency   subject to the power budget at the BS  and the individual ULU and DLU information rate thresholds. The ULU and DLU information rate threshold constraints are crucial to resolve the so-called user fairness since the BS will favor users with a good channel condition. However, such additional rate threshold constraints were not addressed in \cite{AhnTWC16,Dan:TWC:14}.
The  residual SI and  CCI  are also taken into account, which potentially results in a practical system but leads to  more challenging optimizations.

To the best of the authors' knowledge, existing works have not addressed the present optimization problem,
and it  is difficult to even find a feasible point because the feasible set is nonconvex and disconnected. 
 In fact,  the  problem under consideration is very complicated, and its objective function is also highly nonlinear in the involved variables,
for which the optimal solutions are  computationally difficult. Nevertheless, we
develop an iterative algorithm to directly handle the nonconvexity of the considered problem. Our main contributions are summarized as follows:
\begin{itemize}
   \item We propose a new grouping method  to optimize simultaneous uplink  and downlink information transmissions
    by exploring the FD radio at the BS for each group.
\item We first develop an iterative, low-complexity algorithm to obtain the computational solution of  the downlink beamformers and uplink transmit power allocation.  Here  we completely avoid rank-one constraints, which helps reduce the total dimensions of the beamformer vector variables compared to solving the covariance matrices \cite{Dan:TWC:14}.
   \item Since the joint optimization problem of all involved variables consists of  a nonconvex mixed-integer program and each user is served in multiple groups, the gradient projection method in \cite{SanjabiTSP14} is not applicable. Thus, we develop a novel iterative algorithm that arrives at a convex quadratic program at each iteration.
	\item  The obtained solutions are at least local optima since they satisfy the Karush-Kuhn-Tucker (KKT) conditions. Numerical results are presented to confirm the novelty of the proposed algorithms, and the results show that the proposed algorithm converges quite fast and significantly improves the system
performance over conventional FD and HD systems.

\end{itemize}
\subsection{Paper Organization and Notation}
The rest of this paper is organized as follows.  The system model and  problem formulation for the SRM are described in Section~\ref{System Model}. We devise  the optimal solution to the SRM problem for a joint beamformer design and power allocation in Section \ref{sec:PowerAllocation}. The optimal solution for whole problem is presented in Section~\ref{Sec:TimeGroup}. Numerical results are provided  in Section \ref{NumericalResults}, and Section \ref{Conclusion} concludes the paper.

Bold lower and upper case letters respectively represent vectors and matrices. $\mathbf{X}^{H}$, $\mathbf{X}^{T}$,  and $\tr(\mathbf{X})$  are the Hermitian transpose, normal transpose,   and trace of a matrix $\mathbf{X}$, respectively. $\|\cdot\|$ and $|\cdot|$ denote the Euclidean norm of a matrix or vector and the absolute value of a complex scalar, respectively. $\mathbf{I}_N$ represents an $N\times N$ identity matrix. $\mathbf{x}\sim\mathcal{CN}(\boldsymbol{\eta},\boldsymbol{Z})$ means that $\mathbf{x}$ is a random vector following a complex circular Gaussian distribution with mean vector $\boldsymbol{\eta}$ and covariance matrix $\boldsymbol{Z}$. $\mathbb{E}[\cdot]$ denotes the statistical expectation. The notation $\mathbf{X}\succeq\mathbf{0}$ ($\mathbf{X}\succ\mathbf{0}$) means the matrix
$\mathbf{X}$ is positive semi-definite (definite).  $\Re\{\cdot\}$ represents real part of the argument.

\section{System Model and Optimization Problem Formulation} \label{System Model}

\subsection{Signal Model}

\begin{figure}[t]
\centering
\includegraphics[width=0.5\textwidth,trim={-0cm 0.0cm 0cm 2.7cm}]{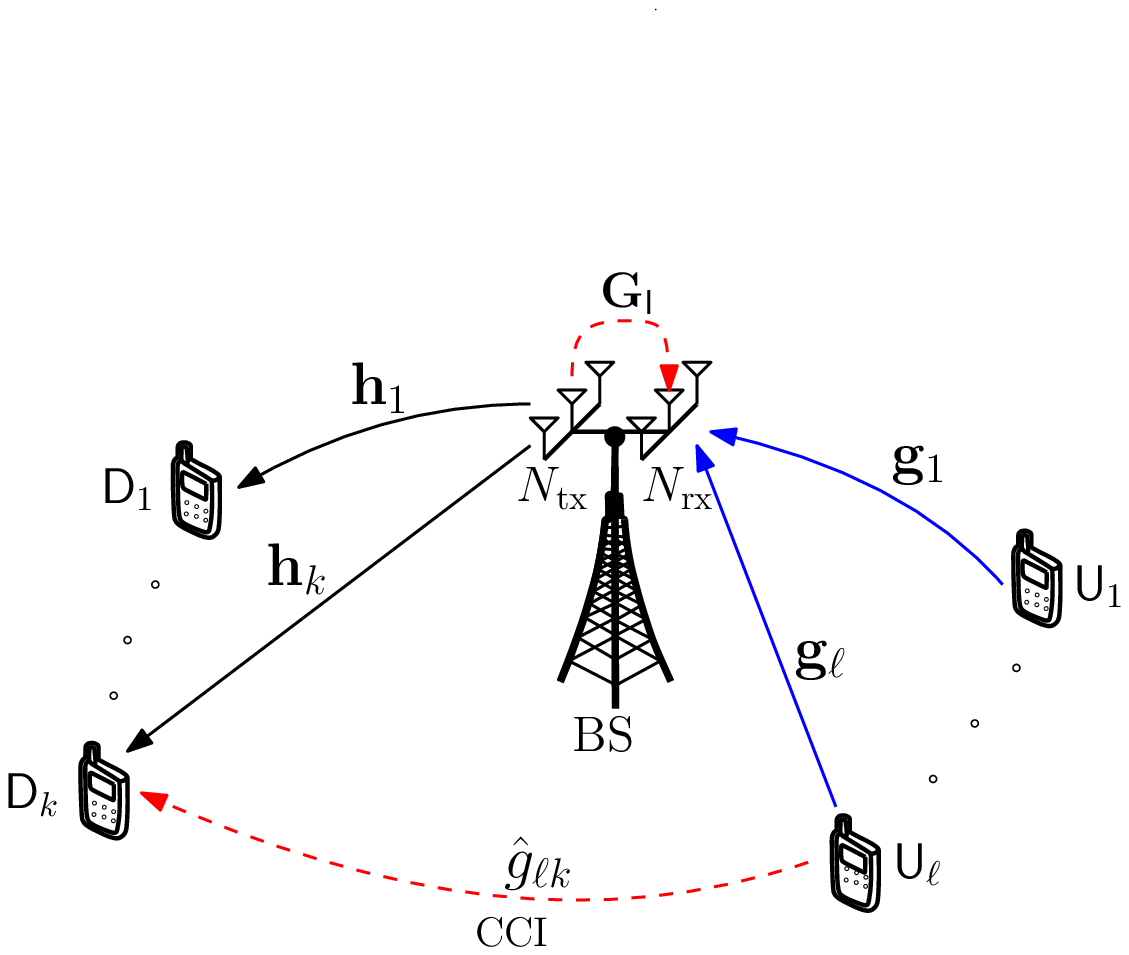}
\caption{A multiuser system model with  FD-enabled BS. The solid and dashed lines denote the transmission  and interference links, respectively.}
\label{fig:SM:1}
\end{figure}

\begin{figure}[t]
\centering
\includegraphics[width=0.48\textwidth]{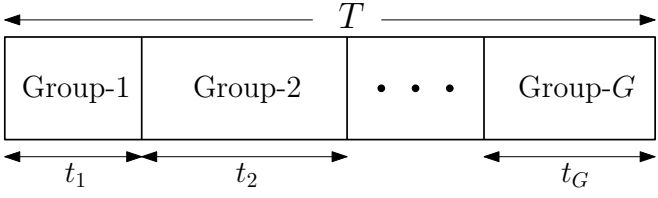}
\caption{Time allocation for different groups.}
\label{fig:TDMA}
\end{figure}

We consider the multiuser system illustrated in Fig.~\ref{fig:SM:1}, which consists of an FD-enabled BS,
$K$ DLUs and $L$ ULUs. The BS is equipped with $N_{\mathrm{rx}}$ receive antennas  and $N_{\mathrm{tx}}$ transmit antennas, and each user
is equipped with a single antenna. Let us define $\mathcal{K}$ and $\mathcal{L}$ to be the sets of all DLUs and ULUs, i.e., $\mathcal{K}\triangleq\{1,2,\cdots,K\}$ and $\mathcal{L}\triangleq\{1,2,\cdots,L\}$, respectively. All channels are assumed to follow independent quasi-static flat fading, i.e., remaining   constant during a communication time block, $T$, but changing independently from one block to another. The channel
state information (CSI) is assumed perfectly available at both the BS and users since it is easily obtained by requesting all DLUs
and ULUs to send their pilots to the BS. In addition, the results under perfect CSI may act as an upper bound on the SR performance for the FD systems. We  assume that all DLUs and ULUs are divided into $G\, (G > 1)$ groups. As shown in Fig.~\ref{fig:TDMA}, the users in each group are served in a separate time slot $g\in\mathcal{G} \triangleq \{1,2,\cdots,G\}$,  and they cause no interference to users in other groups. Each group  still operates in the FD mode. Throughout the paper, $\mathsf{D}_k$ and $\mathsf{U}_{\ell}$ refer to the $k$-th and $\ell$-th user in the downlink and uplink channels, respectively.

 The DL signals are precoded at the BS prior to being transmitted to the DLUs. Then, the received signal at DLU $\mathsf{D}_k$ in the $g$-th group (time slot) and at the BS can be written as
\begin{IEEEeqnarray}{rCl}
y_{\mathsf{D}_k}^g&=&\mathbf{h}_{k}^H\mathbf{w}_{k}^gx_k^g+\sum_{i=1, i\neq k}^K\mathbf{h}_{k}^H\mathbf{w}_{i}^gx_i^g+\sum_{\ell=1}^Lp_{\ell}^g\hat{g}_{\ell k} \tilde{x}_{\ell}^g+n_{\mathsf{D}_k}^g,\nonumber\\
&&\qquad\forall k\in \mathcal{K}, \forall g\in\mathcal{G}, \label{eq:signal:2:a}
\end{IEEEeqnarray}
and
\begin{IEEEeqnarray}{rCl}
\mathbf{y}_{\mathsf{U}}^g&=&\sum_{\ell=1}^L p_{\ell}^g\mathbf{g}_{\ell} \tilde{x}_{\ell}^g+\sqrt{\rho}\sum_{k=1}^K\mathbf{G}_{\mathsf{I}}^H\mathbf{w}_{k}^gx_k^g+\mathbf{n}_{\mathsf{U}}^g,\ \forall g\in\mathcal{G}\label{eq:signal:2:b},
\end{IEEEeqnarray}
respectively. $\mathbf{h}_{k}\in\mathbb{C}^{N_{\mathrm{tx}}\times 1}$,
 $\mathbf{w}_{k}^g\in\mathbb{C}^{N_{\mathrm{tx}}\times 1}$, and $x_k^g$ with $ \mathbb{E}\{|x_k^g|^2\} = 1$ are the transmit channel vector, beamforming vector,  and the message intended for the DLU $\mathsf{D}_k$, respectively, in the $g$-th group. $p_{\ell}^g\in\mathbb{C}$, $\mathbf{g}_{{\ell}}\in\mathbb{C}^{N_{\mathrm{rx}}\times 1}$, and $\tilde{x}_{\ell}^g$ with $ \mathbb{E}\{|\tilde{x}_{\ell}^g|^2\} = 1$ are the transmit power,  receive channel vector, and  message  of ULU $\mathsf{U}_{\ell}$, respectively,  in the $g$-th group. $n_{\mathsf{D}_k}^g\sim\mathcal{CN}(0,\sigma_k^2)$ and $\mathbf{n}_{\mathsf{U}}^g\sim\mathcal{CN}(\mathbf{0},\sigma^2\mathbf{I})$
denote the additive white Gaussian noise (AWGN) at DLU $\mathsf{D}_k$ and at the BS, respectively,  in the $g$-th group.
We assume that the receive AWGNs at each user and  BS are independent of each other.
The term $\sqrt{\rho}\sum_{k=1}^K\mathbf{G}_{\mathsf{I}}^H\mathbf{w}_{k}^g x_k^g$ in (\ref{eq:signal:2:b})
represents the residual SI  after all real-time cancellation in analog and digital domains
\cite{Saetal14}, where $\mathbf{G}_{\mathsf{I}}\in\mathbb{C}^{N_{\mathrm{tx}}\times N_{\mathrm{rx}}}$ is a fading loop channel from the transmit antennas to the receive antennas at the BS and
$0\leq\rho\leq 1$ is used to model the degree of  SI propagation \cite{Riihonen-SP-11}. The term $\sum_{\ell=1}^Lp_{\ell}^g\hat{g}_{\ell k} \tilde{x}_{\ell}^g$ represents the CCI from  ULUs   to DLUs, where $\hat{g}_{\ell k}$ is the complex channel coefficient from $\mathsf{U}_{\ell}$ to $\mathsf{D}_k$.

We assume that the signals of each user in different group/time slot are independent. From \eqref{eq:signal:2:a}, the signal-to-interference-plus-noise ratio (SINR) at the DLU $\mathsf{D}_k$ in the $g$-th group can be expressed as
\begin{equation}\label{eq:SINR:DL:2}
\gamma_{k}^g(\mathbf{w},\mathbf{p})=
\frac{|\mathbf{h}_{k}^H\mathbf{w}_{k}^g|^2}{\sum_{i=1, i\neq k}^K|\mathbf{h}_{k}^H\mathbf{w}_{i}^g|^2+\sum_{\ell=1}^L (p_{\ell}^g)^2|\hat{g}_{\ell k}|^2+\sigma^2_k},
\end{equation}
where $\mathbf{w}\triangleq[\mathbf{w}_k^g]_{k\in\mathcal{K}, g\in\mathcal{G}}$ and $\mathbf{p}\triangleq[p_{\ell}^g]_{\ell\in\mathcal{L},g\in\mathcal{G}}$ are the vectors encompassing the beamformers of all DLUs and  the transmit powers of all ULUs, respectively.
For ULUs, we adopt the minimum mean square error and successive interference cancellation (MMSE-SIC)  receiver at the BS to maximize the received SINR of $\mathsf{U}_{\ell}$ in  \eqref{eq:signal:2:b}.  For simplicity, we assume that the decoding order follows the ULU index, i.e., $\ell =1,2,\cdots, L$. Thus,
the resulting SINR in decoding  $\mathsf{U}_{\ell}$'s information in the $g$-th group can be expressed as \cite{Tse:book:05}
\begin{IEEEeqnarray}{rCl}\label{eq:SINR:UL:2}
\gamma_{\ell}^g(\mathbf{w},\mathbf{p})&=&\,(p_{\ell}^g)^2\mathbf{g}_{\ell}^H\Bigl(\sum_{j>\ell}^L(p^g_{j})^2\mathbf{g}_{j}\mathbf{g}_{j}^H\nonumber\\
&&\quad +\rho\sum_{k=1}^K\mathbf{G}_{\mathsf{I}}^H\mathbf{w}_{k}^g(\mathbf{w}_{k}^g)^H\mathbf{G}_{\mathsf{I}}+
\sigma^2\mathbf{I}\Bigr)^{-1}\mathbf{g}_{\ell}.
\end{IEEEeqnarray}

Let $\alpha_k^g\in\{0,1\}$ and $\beta_{\ell}^g\in\{0,1\}$ be  the binary variables indicating the association of $\mathsf{D}_k$ and $\mathsf{U}_{\ell}$ to the $g$-th group, respectively, i.e.,
\begin{IEEEeqnarray}{rCl}
\alpha_k^g\, (\beta_{\ell}^g)  = \left\{
							\begin{array}{ll}
									1, \mbox{ if  $\mathsf{D}_k$ $(\mathsf{U}_{\ell})$ is served in the $g$-th group},   \\
									0, \mbox{ otherwise}.
							\end{array}
			   \right.
\label{eq:proba}\quad
\end{IEEEeqnarray}
From \eqref{eq:SINR:DL:2}, \eqref{eq:SINR:UL:2}, and \eqref{eq:proba}, 
the  rates of $\mathsf{D}_k$ and $\mathsf{U}_{\ell}$  in the $g$-th group can be found as
\begin{equation}\label{eq:DL:rate}
R_{k}^g\bigr(\mathbf{w},  \mathbf{p}, \boldsymbol{\alpha}\bigr)=\alpha_k^g\ln\Bigl(1+\gamma_{k}^g\bigr(\mathbf{w}, \mathbf{p}\bigr)\Bigr),
\end{equation}
and
\begin{equation}\label{eq:UL:rate}
R_{\ell}^g\bigr(\mathbf{w},  \mathbf{p}, \boldsymbol{\beta}\bigr) = \beta_{\ell}^g\ln\Bigl(1+\gamma_{\ell}^g\bigr(\mathbf{w},\mathbf{p}\bigr)\Bigr),
\end{equation}
where $\boldsymbol{\alpha}\triangleq [\alpha_k^g]_{k\in\mathcal{K},g\in\mathcal{G}}$ and $\boldsymbol{\beta}\triangleq [\beta_{\ell}^g]_{\ell\in\mathcal{L},g\in\mathcal{G}}$.

Let $t_g, g=1,2,\cdots, G$, be the  fraction of time block $T$  allocated for the users in the $g$-th group as shown in Fig.~\ref{fig:TDMA}. Without loss of generality, the time block $T$ is normalized to 1 and then $\sum_{g=1}^{G}t_g \leq 1$. The  achieved rates of $\mathsf{D}_k$ and $\mathsf{U}_{\ell}$ summed over $G$ groups are given as \cite{RazaviyaynTSP14}
\begin{IEEEeqnarray}{rCl}\label{eq:ULDL:rate1}
 R_{k}\bigr(\mathbf{w},  \mathbf{p}, \boldsymbol{\alpha},\boldsymbol{t}\bigr)&=&\sum_{g=1}^{G}t_gR_{k}^g\bigr(\mathbf{w},  \mathbf{p}, \boldsymbol{\alpha}\bigr),
\end{IEEEeqnarray}
and
\begin{IEEEeqnarray}{rCl}\label{eq:ULDL:rate2}
 R_{\ell}\bigr(\mathbf{w},  \mathbf{p}, \boldsymbol{\beta},\boldsymbol{t}\bigr)&=&\sum_{g=1}^{G}t_gR_{\ell}^g\bigr(\mathbf{w},  \mathbf{p}, \boldsymbol{\beta}\bigr), 
\end{IEEEeqnarray}
where $\boldsymbol{t}\triangleq [t_g]_{g\in\mathcal{G}}$.
\begin{remark}
We note that the proposed grouping method does not require additional  communication time block since the total time block for the uplink and downlink transmissions is the same as in \cite{Dan:TWC:14}.
\end{remark}
\subsection{Optimization Problem Formulation}
The aim of this paper is to maximize  the total  SR  of the system by jointly optimizing
 the time allocation, user grouping, DL beamformers, and  UL transmit power allocation  under the transmit power constraints.
In particular, we consider the following optimization problem:
\begin{IEEEeqnarray}{rCl}\label{eq:problem:1}
\underset{\mathbf{w},  \mathbf{p}, \boldsymbol{\alpha}, \boldsymbol{\beta},\boldsymbol{t}}{\mathrm{maximize}}\ && \sum_{k=1}^K\sum_{g=1}^{G}t_gR_{k}^g\bigr(\mathbf{w},  \mathbf{p}, \boldsymbol{\alpha}\bigr) \nonumber\\
&&\qquad + \sum_{\ell =1}^{L}\sum_{g=1}^{G}t_gR_{\ell}^g\bigr(\mathbf{w},  \mathbf{p}, \boldsymbol{\beta}\bigr)\IEEEyessubnumber \label{eq:modified:a}\\
                          \st \;&& \sum\nolimits_{g=1}^{G}t_gR_{k}^g\bigr(\mathbf{w},  \mathbf{p}, \boldsymbol{\alpha}\bigr) \geq \bar{R}^{\mathsf{D}}_k,\;\forall k \in\mathcal{K},\IEEEyessubnumber \label{eq:modified:h} \qquad\\
							&& \sum\nolimits_{g=1}^{G}t_gR_{\ell}^g\bigr(\mathbf{w},  \mathbf{p}, \boldsymbol{\beta}\bigr) \geq \bar{R}^{\mathsf{U}}_\ell,\; \forall \ell \in\mathcal{L},\IEEEyessubnumber \label{eq:modified:i}\\
													&&\sum\nolimits_{k=1}^K\sum\nolimits_{g=1}^Gt_g\|\mathbf{w}_k^g\|^2 \leq P_{bs},\IEEEyessubnumber \label{eq:modified:b}\\
							&&  \sum\nolimits_{g=1}^{G}t_g(p^g_{\ell})^2\leq P_{\ell},\;\forall \ell \in\mathcal{L},\IEEEyessubnumber \label{eq:modified:c}\\
							&&  p_{\ell}^g\geq 0,\; \forall \ell \in\mathcal{L},\; \forall g\in\mathcal{G},   \IEEEyessubnumber \label{eq:modified:pd}\\
							&&\alpha_k^g\in\{0,1\},\; \forall k\in\mathcal{K},\; \forall g\in\mathcal{G},\IEEEyessubnumber \label{eq:modified:e}\\		
							&&		\beta_{\ell}^g\in\{0,1\},\; \forall \ell\in\mathcal{L},\; \forall g\in\mathcal{G},\IEEEyessubnumber \label{eq:modified:f}\\
							&&		\sum\nolimits_{g=1}^G t_g \leq 1\;\text{and}\; t_g \geq 0,\; \forall g\in\mathcal{G},\quad\IEEEyessubnumber \label{eq:modified:g} 
\end{IEEEeqnarray}
where $ \bar{R}^{\mathsf{D}}_k $ and $ \bar{R}^{\mathsf{U}}_\ell $ are the predetermined rate thresholds of each individual DLU and ULU, respectively. $P_{bs}$ and $P_{\ell}$ are the transmit power budgets at the BS and  ULU $\mathsf{U}_{\ell}$, respectively. Note that the different types of constraints corresponding to \eqref{eq:modified:b} and \eqref{eq:modified:c} are  \cite{RazaviyaynTSP14}
\begin{IEEEeqnarray}{rCl}\label{eq:PT:conv:BS}
&&\sum_{k=1}^K\sum_{g=1}^G\|\mathbf{w}_k^g\|^2 \leq P_{bs},\IEEEyessubnumber\label{eq:PT:conv:BSa}\\
&&\sum_{g=1}^{G}(p^g_{\ell})^2 \leq P_{\ell},\;\forall \ell \in\mathcal{L}.\IEEEyessubnumber\label{eq:PT:conv:BSb}
\end{IEEEeqnarray}
However, with the constraints in \eqref{eq:PT:conv:BS}, the BS  and ULUs do not use all allowable power since \eqref{eq:PT:conv:BS}  are just a sum of averaged powers. As a result, the achieved SR of the system may not be optimal. 

We observe that the optimization problem \eqref{eq:problem:1} is NP-hard \cite{Aardal:02} since it is a nonconvex mixed integer program. Therefore, to facilitate the optimization, we first consider a fixed value of  $(\boldsymbol{\alpha}, \boldsymbol{\beta},\boldsymbol{t})$ and develop an iterative algorithm to obtain a locally optimal solution for $(\mathbf{w},  \mathbf{p})$ in Section~\ref{sec:PowerAllocation}. Then, in Section~\ref{Sec:TimeGroup}, we present a joint optimization approach by considering  $(\boldsymbol{\alpha}, \boldsymbol{\beta},\boldsymbol{t})$ as the optimization variables.

\section{Joint Beamformer Design and Power Allocation}\label{sec:PowerAllocation}

\subsection{Proposed Low-Complexity Algorithm}

In this section, we first develop a numerical method to solve \eqref{eq:problem:1} for a fixed value of $(\boldsymbol{\alpha}, \boldsymbol{\beta},\boldsymbol{t})$. Note that the major complexity of solving \eqref{eq:problem:1} is to find the optimal solution for $(\mathbf{w},  \mathbf{p})$. For simplicity, $(\boldsymbol{\alpha}, \boldsymbol{\beta},\boldsymbol{t})$ are set to  $\alpha_{k}^g = \beta_{\ell}^g=1, \forall k,\ell, g$ and $t_g = 1/G,\ \forall g$. This implies that each user is served in $G$ groups and each group is allocated an equal fraction of time. Consequently, the optimization problem can be restated as
\begin{IEEEeqnarray}{rCl}\label{eq:problem:2}
\underset{\mathbf{w},\;  \mathbf{p}}{\mathrm{maximize}}\ && \sum_{k=1}^K\sum_{g=1}^{G}t_g\alpha_k^g\ln\Bigl(1+\gamma_{k}^g\bigr(\mathbf{w}, \mathbf{p}\bigr)\Bigr) \nonumber\\
&&\qquad + \sum_{\ell =1}^{L}\sum_{g=1}^{G}t_g\beta_{\ell}^g\ln\Bigl(1+\gamma_{\ell}^g\bigr(\mathbf{w},\mathbf{p}\bigr)\Bigr)\IEEEyessubnumber \label{eq:problem:a}\\
                          \st \;&&\sum_{g=1}^{G}t_g\alpha_k^g\ln\Bigl(1+\gamma_{k}^g\bigr(\mathbf{w}, \mathbf{p}\bigr)\Bigr) \geq \bar{R}^{\mathsf{D}}_k,\forall k \in\mathcal{K},\qquad\IEEEyessubnumber \label{eq:problem:e} \ \\
						&& \sum_{g=1}^{G}t_g\beta_{\ell}^g\ln\Bigl(1+\gamma_{\ell}^g\bigr(\mathbf{w},\mathbf{p}\bigr)\Bigr) \geq \bar{R}^{\mathsf{U}}_\ell, \forall \ell \in\mathcal{L},\IEEEyessubnumber \label{eq:problem:f}\\
													&&\sum_{k=1}^K\sum_{g=1}^Gt_g\|\mathbf{w}_k^g\|^2 \leq P_{bs},\IEEEyessubnumber \label{eq:problem:b}\\
						&&  \sum_{g=1}^{G}t_g(p^g_{\ell})^2\leq P_{\ell},\;\forall \ell \in\mathcal{L},\IEEEyessubnumber \label{eq:problem:c}\\
						&&  p_{\ell}^g\geq 0,\; \forall \ell \in\mathcal{L}, \forall g\in\mathcal{G}.   \IEEEyessubnumber \label{eq:problem:pd} 
\end{IEEEeqnarray}
After solving \eqref{eq:problem:2}, the ULUs and DLUs can be grouped by checking the optimal values ($\mathbf{w}^*,\mathbf{p}^*$), i.e.,
\begin{equation}\nonumber
\alpha_k^g  = \left\{
							\begin{array}{ll}
									1, \mbox{ if  $\|\mathbf{w}_k^{g,*}\| > \epsilon$},  \forall k\in \mathcal{K}, \forall g\in\mathcal{G}, \\
									0, \mbox{ if  $\|\mathbf{w}_k^{g,*}\| \leq \epsilon$}, \forall k\in \mathcal{K}, \forall g\in\mathcal{G},
							\end{array}
			   \right.
\label{eq:checkingpowersDL}
\end{equation}
and 
\begin{equation}\nonumber
\beta_{\ell}^g  = \left\{
							\begin{array}{ll}
									1, \mbox{ if  $ |p_{\ell}^{g,*}| > \epsilon$},  \forall \ell\in \mathcal{L}, \forall g\in\mathcal{G}, \\
									0, \mbox{ if  $ |p_{\ell}^{g,*}| \leq \epsilon$},  \forall \ell\in \mathcal{L}, \forall g\in\mathcal{G}.
							\end{array}
			   \right.
\label{eq:checkingpowersUL}
\end{equation}
where $\epsilon$ (close to 0) is a predetermined threshold.

Finding an optimal solution to  \eqref{eq:problem:2} is challenging due to the non-concavity of its objective function.  In what follows, we propose an iterative algorithm to obtain a local optimum. We first observe that the constraints \eqref{eq:problem:b}-\eqref{eq:problem:pd} are convex for a fixed value of $(\boldsymbol{\alpha}, \boldsymbol{\beta},\boldsymbol{t})$, while the constraints \eqref{eq:problem:e} and \eqref{eq:problem:f} are nonconvex  and the objective function is also nonconcave. 

Let us treat $\ln\bigl(1+\gamma_{k}^g\bigr(\mathbf{w}, \mathbf{p}\bigr)\bigr)$  first. In the spirit of \cite{WES06},  we express $\ln\bigl(1+\gamma_{k}^g\bigr(\mathbf{w}, \mathbf{p}\bigr)\bigr)$ as
\begin{IEEEeqnarray}{rCl}\label{eq:10a:1}
\ln\Bigl(1+\gamma_{k}^g\bigr(\mathbf{w}, \mathbf{p}\bigr)\Bigr) &\geq& \ln\Biggl(1+\frac{\bigl(\Re\{\mathbf{h}_{k}^H\mathbf{w}_{k}^g\}\bigr)^2}{(\phi_k^g)^2}\Biggr),\nonumber\\
&&\qquad \forall k\in\mathcal{K}, \forall g\in\mathcal{G},\label{eq:10a:1a}
\end{IEEEeqnarray}
with the additional convex constraints
\begin{IEEEeqnarray}{rCl}\label{eq:10a:common}
&&\Bigr(\sum_{i=1, i\neq k}^K|\mathbf{h}_{k}^H\mathbf{w}_{i}^g|^2+\sum_{\ell=1}^L (p_{\ell}^g)^2|\hat{g}_{\ell k}|^2+\sigma^2_k\Bigr)^{1/2} \leq \phi_k^g,\nonumber\\ 
&&\qquad\qquad\qquad\quad\forall k\in\mathcal{K}, \forall g\in\mathcal{G}\IEEEyessubnumber \label{eq:10a:1b}, \\
&&\Re\{\mathbf{h}_{k}^H\mathbf{w}_{k}^g\} \geq 0,\ \forall k\in\mathcal{K}, \forall g\in\mathcal{G}\IEEEyessubnumber \label{eq:10a:1c},
\end{IEEEeqnarray}
where the new optimization variables $\boldsymbol{\phi} \triangleq [\phi_k^g]_{k\in\mathcal{K},g\in\mathcal{G}}$ represent the inter-user interference plus noise experienced by DLU $\mathsf{D}_k$. The equivalence of \eqref{eq:DL:rate} and \eqref{eq:10a:1}  can be easily recognized by
noting  that  constraint \eqref{eq:10a:1b} holds with equality at optimum for a given $\alpha_k^g$. Since the constraints \eqref{eq:10a:1b} and \eqref{eq:10a:1c} are convex, we now only deal with the non-concave function \eqref{eq:10a:1a}. From \eqref{eq:10a:1a}, it follows that 
\begin{IEEEeqnarray}{rCl}\label{eq:10a:2}
\ln\Bigl(1+\gamma_{k}^g\bigr(\mathbf{w}, \mathbf{p}\bigr)\Bigr) \geq - \ln\Biggl(1-\frac{\bigl(\Re\{\mathbf{h}_{k}^H\mathbf{w}_{k}^g\}\bigr)^2}{(\phi_k^g)^2 + \bigl(\Re\{\mathbf{h}_{k}^H\mathbf{w}_{k}^g\}\bigr)^2}\Biggr). \quad\;\;
\end{IEEEeqnarray}
It is obvious that $0 \leq\frac{\bigl(\Re\{\mathbf{h}_{k}^H\mathbf{w}_{k}^g\}\bigr)^2}{(\phi_k^g)^2 + \bigl(\Re\{\mathbf{h}_{k}^H\mathbf{w}_{k}^g\}\bigr)^2} < 1$, thus the right-hand side (RHS) of \eqref{eq:10a:2} is a convex function with respect to ($\mathbf{w}, \boldsymbol{\phi}$) \cite{Stephen}. At this point,  we apply  an inner approximation convex method \cite{Marks:78} for \eqref{eq:10a:2}. Let us define a feasible point for $x$ at the $(n + 1)$-th iteration in an iterative algorithm presented shortly as denoted by $x^{(n)}$.  At the feasible point $(\mathbf{w}^{(n)}, \boldsymbol{\phi}^{(n)})$, a global lower bound of RHS of \eqref{eq:10a:2} can be obtained as \cite{Tuy-B-00}
\begin{IEEEeqnarray}{rCl}\label{eq:10a:3}
\ln\Bigl(1+\gamma_{k}^g\bigr(\mathbf{w}, \mathbf{p}\bigr)\Bigr) &\geq& \varphi_k^{g,(n)} + \chi_{k}^{g,(n)}\Re\{\mathbf{h}_{k}^H\mathbf{w}_{k}^{g}\} \nonumber\\
&&- \varpi_{k}^{g,(n)}\Bigl((\phi_k^{g})^2 + \bigl(\Re\{\mathbf{h}_{k}^H\mathbf{w}_{k}^{g}\}\bigr)^2\Bigr)\quad\; \label{eq:10a:3a}\\
                 &:=& \mathcal{F}_k^{g,(n)}(\mathbf{w}, \boldsymbol{\phi})\label{eq:10a:3b},
\end{IEEEeqnarray}
where $\varphi_k^{g,(n)}$, $\chi_{k}^{g,(n)}$, and $\varpi_{k}^{g,(n)}$ are defined as
\begin{IEEEeqnarray}{rCl}\label{eq:10a:4}
\varphi_k^{g,(n)} &=& - \ln\Biggl(1-\frac{\bigl(\Re\{\mathbf{h}_{k}^H\mathbf{w}_{k}^{g,(n)}\}\bigr)^2}{(\phi_k^{g,(n)})^2 + \bigl(\Re\{\mathbf{h}_{k}^H\mathbf{w}_{k}^{g,(n)}\}\bigr)^2}\Biggr) \nonumber\\
&&- \frac{\bigl(\Re\{\mathbf{h}_{k}^H\mathbf{w}_{k}^{g,(n)}\}\bigr)^2}{(\phi_k^{g,(n)})^2},\nonumber\\
 \chi_{k}^{g,(n)} &=& 2\frac{\Re\{\mathbf{h}_{k}^H\mathbf{w}_{k}^{g,(n)}\}}{(\phi_k^{g,(n)})^2},\nonumber\\
\varpi_{k}^{g,(n)} &=&\frac{\bigl(\Re\{\mathbf{h}_{k}^H\mathbf{w}_{k}^{g,(n)}\}\bigr)^2}{(\phi_k^{g,(n)})^2\Bigl((\phi_k^{g,(n)})^2 + \bigl(\Re\{\mathbf{h}_{k}^H\mathbf{w}_{k}^{g,(n)}\}\bigr)^2\Bigr)}.
\end{IEEEeqnarray}
It should be noted that $\mathcal{F}_k^{g,(n)}(\mathbf{w}, \boldsymbol{\phi})$ is concave and \eqref{eq:10a:3} is active at  optimum, i.e.,
\begin{IEEEeqnarray}{rCl}\label{eq:10a:5}
\mathcal{F}_k^{g,(n)}(\mathbf{w}^{(n)}, \boldsymbol{\phi}^{(n)}) = \ln\Bigl(1+\gamma_{k}^g\bigr(\mathbf{w}^{(n)}, \mathbf{p}^{(n)}\bigr)\Bigr). 
\end{IEEEeqnarray}
Clearly, $\ln\bigl(1+\gamma_{k}^g(\mathbf{w}, \mathbf{p})\bigr)$ can be iteratively replaced by  $\mathcal{F}_k^{g,(n)}(\mathbf{w}, \boldsymbol{\phi})$ to achieve a  local optimum \cite{Marks:78}.

We now turn our attention to $\ln\bigl(1+\gamma_{\ell}^g(\mathbf{w},\mathbf{p})\bigr)$. The epigraph of $\gamma_{\ell}^g(\mathbf{w},\mathbf{p})$ can be written as $\{(z_{\ell}^g,\mathbf{w},\mathbf{p})| z_{\ell}^g \geq \gamma_{\ell}^g(\mathbf{w},\mathbf{p})\}$, which is equivalent to 
\begin{equation}\label{eq:10a:6}
\begin{bmatrix}
   z_{\ell}^g       & p_{\ell}^g\mathbf{g}_{\ell}^H\\
    p_{\ell}^g\mathbf{g}_{\ell}       & \boldsymbol{\Xi}(\mathbf{w},\mathbf{p})
\end{bmatrix}\succeq \mathbf{0},
\end{equation}
where  $\boldsymbol{\Xi}(\mathbf{w},\mathbf{p})\triangleq\sum_{j>\ell}^L(p^g_{j})^2\mathbf{g}_{j}\mathbf{g}_{j}^H+
\rho\sum_{k=1}^K\mathbf{G}_{\mathsf{I}}^H\mathbf{w}_{k}^g(\mathbf{w}_{k}^g)^H\mathbf{G}_{\mathsf{I}}+
\sigma^2\mathbf{I} $, and \eqref{eq:10a:6} is obtained by using the Schur complement \cite{Stephen,Dan:TWC:14}. Obviously, the linear matrix inequality in \eqref{eq:10a:6} is convex, leading to a convex set of $\gamma_{\ell}^g(\mathbf{w},\mathbf{p})$ that is useful to develop an inner approximation of  $\ln\bigl(1+\gamma_{\ell}^g(\mathbf{w},\mathbf{p})\bigr)$. Similarly to \eqref{eq:10a:3},
 at the feasible point $(\mathbf{w}^{(n)}, \mathbf{p}^{(n)})$, 
$\ln\bigl(1+\gamma_{\ell}^g(\mathbf{w},\mathbf{p})\bigr)$ is lower bounded as \cite{Tuy-B-00}
\begin{IEEEeqnarray}{rCl}\label{eq:10a:7}
\ln\bigl(1+\gamma_{\ell}^g(\mathbf{w},\mathbf{p})\bigr) &\geq& \vartheta_{\ell}^{g,(n)} +   \psi_{\ell}^{g,(n)}p_{\ell}^{g} - \lambda_{\ell}^{g,(n)}(\mathbf{w},\mathbf{p})                      \qquad      \label{eq:10a:7a}\\
&:=&\mathcal{P}_{\ell}^{g,(n)}(\mathbf{w},\mathbf{p}),\label{eq:10a:7b}
\end{IEEEeqnarray}
where $\vartheta_{\ell}^{g,(n)}$, $\psi_{\ell}^{g,(n)}$, and $\lambda_{\ell}^{g,(n)}(\mathbf{w},\mathbf{p}) $ are defined as
\begin{IEEEeqnarray}{rCl}
&&\vartheta_{\ell}^{g,(n)}=\ln\bigl(1+\gamma_{\ell}^g(\mathbf{w}^{g,(n)},\mathbf{p}^{g,(n)})\bigr)
-\gamma_{\ell}^g\bigl(\mathbf{w}^{g,(n)},\mathbf{p}^{g,(n)}\bigr),\nonumber\\
&&\psi_{\ell}^{g,(n)} = 2p_{\ell}^{g,(n)}\mathbf{g}_{\ell}^H\Bigl(\ds\sum_{j>\ell}^L\bigl(p^{g,(n)}_{j}\bigr)^2\mathbf{g}_{j}\mathbf{g}_{j}^H\nonumber\\
&&+\, \rho\sum_{k=1}^K\mathbf{G}_{\mathsf{I}}^H\mathbf{w}^{g,(n)}_{k}\bigl(\mathbf{w}^{g,(n)}_{k}\bigr)^H\mathbf{G}_{\mathsf{I}}+
\sigma^2\mathbf{I}\Bigr)^{-1}\mathbf{g}_{\ell},\nonumber\\
&&\lambda_{\ell}^{g,(n)}(\mathbf{w},\mathbf{p})=(p_{\ell}^g)^2\mathbf{g}_{\ell}^H\boldsymbol{\Theta}^{g,(n)}_{\ell}\mathbf{g}_{\ell}+
\ds\sum_{j>\ell}^L(p_{j}^g)^2\mathbf{g}_{j}^H\boldsymbol{\Theta}^{g,(n)}_{\ell}\mathbf{g}_{j}\nonumber\\
&&\quad\qquad+\ \rho\sum_{k=1}^K(\mathbf{w}_{k}^g)^H\mathbf{G}_{\mathsf{I}}\boldsymbol{\Theta}^{g,(n)}_{\ell}\mathbf{G}_{\mathsf{I}}^H\mathbf{w}_{k}^g+
\sigma^2\tr\bigl(\boldsymbol{\Theta}^{g,(n)}_{\ell}\bigr),\quad \nonumber\\
&&\boldsymbol{\Theta}^{g,(n)}_{\ell}=\Bigl(\ds\sum_{j>\ell}^L\bigl(p^{g,(n)}_{j}\bigr)^2\mathbf{g}_{j}\mathbf{g}_{j}^H+
\rho\sum_{k=1}^K\mathbf{G}_{\mathsf{I}}^H\mathbf{w}^{g,(n)}_{k}\bigl(\mathbf{w}^{g,(n)}_{k}\bigr)^H\mathbf{G}_{\mathsf{I}}\nonumber\\
&&\qquad\quad  +\;  \sigma^2\mathbf{I}\Bigr)^{-1} - \Bigl( \bigl(p^{g,(n)}_{\ell}\bigr)^2\mathbf{g}_{\ell}\mathbf{g}_{\ell}^{H}+ \ds\sum_{j>\ell}^L\bigl(p^{g,(n)}_{j}\bigr)^2\mathbf{g}_{j}\mathbf{g}_{j}^H \nonumber\\
&&\qquad\quad +\; \rho\sum_{k=1}^K\mathbf{G}_{\mathsf{I}}^H\mathbf{w}^{g,(n)}_{k}\bigl(\mathbf{w}^{g,(n)}_{k}\bigr)^H\mathbf{G}_{\mathsf{I}}+
\sigma^2\mathbf{I}\Bigr)^{-1}\succeq \mathbf{0}.\quad
\end{IEEEeqnarray}
We note that $\mathcal{P}_{\ell}^{g,(n)}(\mathbf{w},\mathbf{p})$ is concave and satisfies the following condition:
\begin{IEEEeqnarray}{rCl}\label{eq:10a:9}
\mathcal{P}_{\ell}^{g,(n)}(\mathbf{w}^{(n)},\mathbf{p}^{(n)}) = \ln\bigl(1+\gamma_{\ell}^g(\mathbf{w}^{(n)},\mathbf{p}^{(n)})\bigr).
\end{IEEEeqnarray}

In summary, the approximate convex problem solved at the $(n+1)$-th iteration of the proposed computation procedure is given by
\begin{IEEEeqnarray}{rCl}\label{eq:convexappro:obj}
 \ds\underset{\mathbf{w},\, \mathbf{p},\, \boldsymbol{\phi}}{\mathrm{maximize}}\; &&
\sum_{k=1}^K\sum_{g=1}^{G}t_g\alpha_k^g\mathcal{F}_k^{g,(n)}(\mathbf{w}, \boldsymbol{\phi}) \nonumber \\
&&\qquad\quad +\; \sum_{\ell =1}^{L}\sum_{g=1}^{G}t_g\beta_{\ell}^g\mathcal{P}_{\ell}^{g,(n)}(\mathbf{w},\mathbf{p})    \IEEEyessubnumber \label{eq:convexappro:a}\\
\st\; && \sum_{g=1}^{G}t_g\alpha_k^g\mathcal{F}_k^{g,(n)}(\mathbf{w}, \boldsymbol{\phi}) \geq \bar{R}^{\mathsf{D}}_k,\forall k \in\mathcal{K},\IEEEyessubnumber \label{eq:convexappro:b} \qquad\\
&& \sum_{g=1}^{G}t_g\beta_{\ell}^g\mathcal{P}_{\ell}^{g,(n)}(\mathbf{w},\mathbf{p}) \geq \bar{R}^{\mathsf{U}}_\ell, \forall \ell \in\mathcal{L},\IEEEyessubnumber \label{eq:convexappro:c} \\
&&\eqref{eq:problem:b}, \eqref{eq:problem:c}, \eqref{eq:problem:pd}, \eqref{eq:10a:common}.\IEEEyessubnumber \label{eq:convexappro:d}
\end{IEEEeqnarray}
To solve \eqref{eq:convexappro:obj} by existing solvers, i.e., SDPT3 \cite{Toh} or MOSEK \cite{MOSEK}, we transform \eqref{eq:convexappro:obj} into the following convex program at each iteration:
\begin{IEEEeqnarray}{rCl}\label{eq:convexappro:obj2}
 \ds\underset{\mathbf{w},\, \mathbf{p},\, \boldsymbol{\phi},\boldsymbol{\theta},\tilde{\boldsymbol{\theta}}}{\mathrm{maximize}}\; &&
\sum_{k=1}^K\sum_{g=1}^{G}t_g\alpha_k^g\ddot{\mathcal{F}}_k^{g,(n)}(\mathbf{w}, \boldsymbol{\theta}) \nonumber\\
 && \qquad\quad +\; \sum_{\ell =1}^{L}\sum_{g=1}^{G}t_g\beta_{\ell}^g\ddot{\mathcal{P}}_{\ell}^{g,(n)}(\mathbf{p},\tilde{\boldsymbol{\theta}})    \IEEEyessubnumber \label{eq:convexappro:a2}\\
\st\; &&\sum_{g=1}^{G}t_g\alpha_k^g\ddot{\mathcal{F}}_k^{g,(n)}(\mathbf{w},\boldsymbol{\theta}) \geq \bar{R}^{\mathsf{D}}_k,\forall k \in\mathcal{K},\IEEEyessubnumber \label{eq:convexappro:b11} \\
&& \sum_{g=1}^{G}t_g\beta_{\ell}^g\ddot{\mathcal{P}}_{\ell}^{g,(n)}(\mathbf{p},\tilde{\boldsymbol{\theta}}) \geq \bar{R}^{\mathsf{U}}_\ell, \forall \ell \in\mathcal{L},\IEEEyessubnumber \label{eq:convexappro:c12} \\
&& (\phi_k^{g})^2 + \bigl(\Re\{\mathbf{h}_{k}^H\mathbf{w}_{k}^{g}\}\bigr)^2 \leq \theta_k^g,\forall k \in\mathcal{K},\forall g \in\mathcal{G},\qquad\;\IEEEyessubnumber \label{eq:convexappro:b2} \\
&& \lambda_{\ell}^{g,(n)}(\mathbf{w},\mathbf{p}) \leq \tilde{\theta}_{\ell}^g, \forall \ell \in\mathcal{L},\forall g \in\mathcal{G},\IEEEyessubnumber \label{eq:convexappro:c2} \\
&&\eqref{eq:problem:b}, \eqref{eq:problem:c}, \eqref{eq:problem:pd}, \eqref{eq:10a:common}, \IEEEyessubnumber \label{eq:convexappro:d2}
\end{IEEEeqnarray}
with
\begin{IEEEeqnarray}{rCl}
\ddot{\mathcal{F}}_k^{g,(n)}(\mathbf{w},\boldsymbol{\theta}) &:=& \varphi_k^{g,(n)} + \chi_{k}^{g,(n)}\Re\{\mathbf{h}_{k}^H\mathbf{w}_{k}^{g}\} - \varpi_{k}^{g,(n)}\theta_k^g,\nonumber\\
\ddot{\mathcal{P}}_{\ell}^{g,(n)}(\mathbf{p},\tilde{\boldsymbol{\theta}}) &:=&  \vartheta_{\ell}^{g,(n)} +   \psi_{\ell}^{g,(n)}p_{\ell}^{g} - \tilde{\theta}_{\ell}^g,\nonumber
\end{IEEEeqnarray}
where $\boldsymbol{\theta}\triangleq[\theta_k^g]_{k\in\mathcal{K},g\in\mathcal{G}}$ and $\tilde{\boldsymbol{\theta}}\triangleq[\tilde{\theta}_{\ell}^g]_{\ell\in\mathcal{L},g\in\mathcal{G}}$ are  additional optimization variables to tackle the quadratic functions in \eqref{eq:convexappro:a}.
 By  updating  ($\mathbf{w},\mathbf{p},$$\boldsymbol{\phi}$) for the next iteration, we arrive at the iterative algorithm  to  maximize the total SR of the system, as summarized in Algorithm~\ref{algo:proposed:DUAL}. 

\begin{algorithm}[t]
\begin{algorithmic}[1]

\protect\caption{Proposed iterative algorithm for the SR maximization problem \eqref{eq:problem:2}}

\label{algo:proposed:DUAL}

\global\long\def\algorithmicrequire{\textbf{Initialization:}}

\REQUIRE  Set $n:=0$ and solve \eqref{eq:convexappro:obj22} to generate an initial feasible point $(\mathbf{w}^{(0)},\mathbf{p}^{(0)},\boldsymbol{\phi}^{(0)})$.
\REPEAT
\STATE Solve \eqref{eq:convexappro:obj2} to obtain the optimal solutions $(\mathbf{w}^{*},\mathbf{p}^{*},\boldsymbol{\phi}^{*},\boldsymbol{\theta}^{*},\tilde{\boldsymbol{\theta}}^{*})$.

\STATE Update $\mathbf{w}^{(n+1)}:=\mathbf{w}^{*},$   $\mathbf{p}^{(n+1)}:=\mathbf{p}^{*}$, and
   $ \boldsymbol{\phi}^{(n+1)}:=\boldsymbol{\phi}^{*}$.
\STATE Set $n:=n+1.$
\UNTIL Convergence\\
\end{algorithmic} \end{algorithm}

\textit{Generation of initial points}: Note that if $(\mathbf{w}^{(n)},\mathbf{p}^{(n)},\boldsymbol{\phi}^{(n)})$ are feasible for \eqref{eq:convexappro:obj2}, then the subsequent problems at the $(n+1)$-th iteration are also feasible. Thus,  it is desirable to generate initial values for $(\mathbf{w}^{(0)},\mathbf{p}^{(0)},\boldsymbol{\phi}^{(0)})$ to make sure that Algorithm~\ref{algo:proposed:DUAL} can be efficiently solved  in the first iteration. 
To find a feasible point for the constraints in \eqref{eq:problem:2}, initialized by any feasible $(\mathbf{w}^{(0)},\mathbf{p}^{(0)},\boldsymbol{\phi}^{(0)})$, we successively solve 
\begin{IEEEeqnarray}{rCl}\label{eq:convexappro:obj22}
 \ds\underset{\mathbf{w},\, \mathbf{p},\, \boldsymbol{\phi},\boldsymbol{\theta},\tilde{\boldsymbol{\theta}}}{\mathrm{maximize}}\; && \underset{\substack{k\in\mathcal{K}\\ \ell\in\mathcal{L}}}{\min}\;\Biggl\{\frac{\sum_{g=1}^{G}t_g\alpha_k^g\ddot{\mathcal{F}}_k^{g,(n)}(\mathbf{w},\boldsymbol{\theta})}{\bar{R}^{\mathsf{D}}_k}, \nonumber \\
&&\qquad\quad\frac{\sum_{g=1}^{G}t_g\beta_{\ell}^g\ddot{\mathcal{P}}_{\ell}^{g,(n)}(\mathbf{p},\tilde{\boldsymbol{\theta}})}{\bar{R}^{\mathsf{U}}_\ell}    \Biggl\}\IEEEyessubnumber \label{eq:convexappro:a22}\\
\st\; &&\eqref{eq:problem:b}, \eqref{eq:problem:c}, \eqref{eq:problem:pd}, \eqref{eq:10a:common},\eqref{eq:convexappro:b2}, \eqref{eq:convexappro:c2},\IEEEyessubnumber \label{eq:convexappro:d22}
\end{IEEEeqnarray}
until reaching its optimal value of more than or equal to $1$ to satisfy the constraints in \eqref{eq:problem:2}, i.e.,
\begin{equation}\nonumber
\underset{\substack{k\in\mathcal{K}\\ \ell\in\mathcal{L}}}{\min}\;\Biggl\{\frac{\sum_{g=1}^{G}t_g\alpha_k^g\ddot{\mathcal{F}}_k^{g,(n)}(\mathbf{w},\boldsymbol{\theta})}{\bar{R}^{\mathsf{D}}_k},\frac{\sum_{g=1}^{G}t_g\beta_{\ell}^g\ddot{\mathcal{P}}_{\ell}^{g,(n)}(\mathbf{p},\tilde{\boldsymbol{\theta}})}{\bar{R}^{\mathsf{U}}_\ell}    \Biggl\} \geq 1.
\end{equation}

\subsection{Proof of Convergence and Complexity Analysis}

The convergence of the proposed algorithm is guaranteed due to the monotonic increase in the objective function in problem \eqref{eq:problem:2} after each iteration. In particular, the convergence result for Algorithm~\ref{algo:proposed:DUAL} is stated in the following proposition.
\lbn\begin{proposition}\label{prop1} Algorithm~\ref{algo:proposed:DUAL} generates a sequence $\{(\mathbf{w}^{(n)},\mathbf{p}^{(n)},\boldsymbol{\phi}^{(n)})\}$ of improved points of \eqref{eq:convexappro:obj2} and \eqref{eq:problem:2} leading to a non-decreasing sequence of its objective value, which also converges to a KKT  point.
\end{proposition}\lbn

\begin{IEEEproof} See Appendix A.
\end{IEEEproof}\lbn

\textit{Complexity Analysis}:  The proposed iterative algorithm actually has low complexity in  that it only requires solving  simple convex quadratic and linear constraints at each iteration.   Specifically,  the convex problem \eqref{eq:convexappro:obj2} involves  $( N_{\mathrm{tx}}K + 2L + K)G$ scalar real decision variables and $(4KG +3LG+K+2L+1)$  quadratic and linear constraints. Then, in each iteration of Algorithm~\ref{algo:proposed:DUAL}, the computational complexity for solving \eqref{eq:convexappro:obj2} is $\mathcal{O}\Bigl((4KG +3LG+K+2L+1)^{2.5}\bigl((N_{\mathrm{tx}}K + 2L + K)^2G^2 + 4KG +3LG+K+2L+1\bigr) \Bigr)$ \cite{Ben:2001}.

\section{Time Allocation and Dynamic User Grouping Assignment}\label{Sec:TimeGroup}

In this section, we attempt to jointly optimize the fraction of time allocated for each group and user grouping assignment for each user in order to  maximize the total SR of the system, i.e., considering $(\boldsymbol{\alpha}, \boldsymbol{\beta},\boldsymbol{t})$ as the optimization variables. Although problem \eqref{eq:problem:1} is a nonconvex integer program,  its global optimization can be found. Note that the major difficulty in solving \eqref{eq:problem:1} is to find the optimal solutions for $(\boldsymbol{\alpha}, \boldsymbol{\beta})$ since they are discrete variables. It is obvious that the optimal solutions for  $(\alpha_k^g, \beta_{\ell}^g)$ will be either 1 or 0. Unfortunately, once $\alpha_k^g$ and $\beta_{\ell}^g$ are set to 1 or 0, the grouping assignment for each user will be fixed and thus the obtained results cannot be optimal. To circumvent this issue, we relax the constraints \eqref{eq:modified:e} and \eqref{eq:modified:f} into $0 \leq \alpha_k^g \leq 1$ and $0 \leq \beta_{\ell}^g \leq 1$, respectively. With the convex inner approximation approach presented in Section~\ref{sec:PowerAllocation},  the  optimization problem \eqref{eq:problem:1} can be revised to
\begin{IEEEeqnarray}{rCl}\label{eq:problem:3}
\underset{\mathbf{w},  \mathbf{p},\boldsymbol{\phi}, \boldsymbol{\theta}, \tilde{\boldsymbol{\theta}}, \boldsymbol{\alpha}, \boldsymbol{\beta},\boldsymbol{t}}{\mathrm{maximize}}\ &&
\sum_{k=1}^K\sum_{g=1}^{G}t_g\alpha_k^g\ddot{\mathcal{F}}_k^{g,(n)}(\mathbf{w}, \boldsymbol{\theta}) \nonumber \\
&&\qquad + \sum_{\ell =1}^{L}\sum_{g=1}^{G}t_g\beta_{\ell}^g\mathcal{\ddot{P}}_{\ell}^{g,(n)}(\mathbf{p},\tilde{\boldsymbol{\theta}})    \IEEEyessubnumber \label{eq:problem:3:a} \qquad\\
                          \st \;												
						&& 0 \leq \alpha_k^g \leq1,\; \forall k\in\mathcal{K}, \forall g\in\mathcal{G},\IEEEyessubnumber \label{eq:problem:3:b}\\		
						&&	0 \leq	\beta_{\ell}^g\leq 1,\; \forall \ell\in\mathcal{L}, \forall g\in\mathcal{G},\IEEEyessubnumber \label{eq:problem:3:c}\\
						&&		\eqref{eq:modified:b}, \eqref{eq:modified:c}, \eqref{eq:modified:pd}, \eqref{eq:modified:g}, \eqref{eq:10a:common}, \nonumber\\
						&& \eqref{eq:convexappro:b11}, \eqref{eq:convexappro:c12}, \eqref{eq:convexappro:b2}, \eqref{eq:convexappro:c2}.\IEEEyessubnumber \label{eq:problem:3:d}
\end{IEEEeqnarray}

\begin{remark}
The optimization formulation in \eqref{eq:problem:3} shows the fact that each user can be served in multiple groups as long as its objective value increases. This differs from the conventional orthogonal user grouping  based scheduling strategy, in which each user is involved in only one group/time slot.
\end{remark}

Given an optimal solution for $(\mathbf{w},  \mathbf{p},\boldsymbol{\phi}, \boldsymbol{\theta}, \tilde{\boldsymbol{\theta}})$ obtained from Algorithm~\ref{algo:proposed:DUAL}, a simple method for solving \eqref{eq:problem:3} works as follows. We first fix $(\boldsymbol{\alpha}, \boldsymbol{\beta})$ and solve \eqref{eq:problem:3} with respect to $\boldsymbol{t}$. It is a linear program with an optimal solution that can be easily found  using standard convex optimization techniques \cite{Stephen}. We then solve \eqref{eq:problem:3}  with respect to $(\boldsymbol{\alpha}, \boldsymbol{\beta})$ for a fixed $\boldsymbol{t}$ and repeat the procedure until convergence. However, this method often shows slow convergence and is of relatively high complexity since the major complexity comes from solving \eqref{eq:convexappro:obj2} in Algorithm~\ref{algo:proposed:DUAL}. In what follows, we propose an efficient method to find all optimization variables in a single layer, which greatly reduces the complexity. To begin, we first treat the product of the grouping assignment variables and per-user per-group rate in \eqref{eq:problem:3:a} as
\begin{IEEEeqnarray}{rCl} \label{eq:peruser pergroup rate}
	 \alpha_k^g \ddot{\mathcal{F}}_k^{g,(n)}(\mathbf{w}, \boldsymbol{\theta}) &\geq& \bigl(\tau_k^g\bigr)^2,\,\forall k\in\mathcal{K},\forall g\in\mathcal{G},\IEEEyessubnumber \label{eq:peruser pergroup downlink rate} \\
 \beta_\ell^g \ddot{\mathcal{P}}_\ell^{g,(n)}(\mathbf{p},\tilde{\boldsymbol{\theta}}) &\geq& \bigl(\kappa_\ell^g\bigr)^2, \,\forall \ell\in\mathcal{L},\forall g\in\mathcal{G},\IEEEyessubnumber\label{eq:peruser pergroup uplink rate}
\end{IEEEeqnarray}
where   $ \boldsymbol{\tau} \triangleq \bigl[\tau_k^g\bigr]_{k\in\mathcal{K},g\in\mathcal{G}} $ and $ \boldsymbol{\kappa} \triangleq \bigl[\kappa_\ell^g\bigr]_{\ell\in\mathcal{L},g\in\mathcal{G}} $ are the new optimization variables.  It is clear that the constraints in \eqref{eq:peruser pergroup rate} are convex quadratic functions \cite{Stephen}. In order to jointly optimize $ t_g $ as the optimization variable, we  approximate the RHSs of \eqref{eq:peruser pergroup downlink rate} and \eqref{eq:peruser pergroup uplink rate} at the feasible point $(\tau_k^{g,(n)},\kappa_{\ell}^{g,(n)})$, and the corresponding approximations are transformed to the following linear constraints:  
\begin{IEEEeqnarray}{rCl} \label{eq:peruser pergroup rate approx.}
	&& \bigl(\tau_k^{g,(n)}\bigr)^2 + 2\tau_k^{g,(n)}(\tau_k^{g}-\tau_k^{g,(n)})\geq \hat{\tau}_k^g \geq 0,\,\forall k,\forall g,\IEEEyessubnumber \label{eq:peruser pergroup downlink rate approx.} \qquad\\
	&& \bigl(\kappa_\ell^{g,(n)}\bigr)^2 + 2\kappa_\ell^{g,(n)}(\kappa_\ell^{g}-\kappa_\ell^{g,(n)})\geq \hat{\kappa}_\ell^g \geq 0,\,\forall \ell,\forall g,\quad \IEEEyessubnumber \label{eq:peruser pergroup uplink rate approx.}
\end{IEEEeqnarray}
where  $ \boldsymbol{\hat{\tau}} \triangleq \bigl[\hat{\tau}_k^g\bigr]_{k\in\mathcal{K},g\in\mathcal{G}} $ and $ \boldsymbol{\hat{\kappa}} \triangleq \bigl[\hat{\kappa}_\ell^g\bigr]_{\ell\in\mathcal{L},g\in\mathcal{G}} $  are the new optimization variables.  

Consequently, the optimization problem \eqref{eq:problem:3} can be rewritten as
\begin{IEEEeqnarray}{rCl}\label{eq:problem:31}
\underset{\substack{\mathbf{w},  \mathbf{p},\boldsymbol{\phi}, \boldsymbol{\theta}, \tilde{\boldsymbol{\theta}}, \boldsymbol{\alpha}, \\ \boldsymbol{\beta},\boldsymbol{t},\boldsymbol{\tau},\hat{\boldsymbol{\tau}},\boldsymbol{\kappa},\hat{\boldsymbol{\kappa}}}}{\mathrm{maximize}}\ &&
\sum_{k=1}^K\sum_{g=1}^{G}t_g\hat{\tau}_k^g + \sum_{\ell =1}^{L}\sum_{g=1}^{G}t_g\hat{\kappa}_{\ell}^g    \IEEEyessubnumber \label{eq:problem:31:a}\\
                          \st \;												
						&& \sum\nolimits_{g=1}^{G}t_g\hat{\tau}_k^g \geq \bar{R}^{\mathsf{D}}_k,\; \forall k\in\mathcal{K}, \IEEEyessubnumber \label{eq:problem:31:b}\\		
						&&	\sum\nolimits_{g=1}^{G}t_g\hat{\kappa}_{\ell}^g\geq \bar{R}^{\mathsf{U}}_\ell,\, \forall \ell \in\mathcal{L}, \IEEEyessubnumber \label{eq:problem:31:c}\\
						&&\sum\nolimits_{k=1}^K\sum\nolimits_{g=1}^Gt_g\|\mathbf{w}_k^g\|^2 \leq P_{bs},\IEEEyessubnumber \label{eq:problem:31:d}\\
							&&  \sum\nolimits_{g=1}^{G}t_g(p^g_{\ell})^2\leq P_{\ell},\;\forall \ell \in\mathcal{L},\IEEEyessubnumber \label{eq:problem:31:e}\\
						&&		\eqref{eq:modified:pd}, \eqref{eq:modified:g}, \eqref{eq:10a:common}, \eqref{eq:convexappro:b2}, \eqref{eq:convexappro:c2}, \nonumber\\
						&& \eqref{eq:problem:3:b},\eqref{eq:problem:3:c},\eqref{eq:peruser pergroup rate},\eqref{eq:peruser pergroup rate approx.}.\IEEEyessubnumber \label{eq:problem:31:f}
\end{IEEEeqnarray}
Even after the above transformations, the optimization problem  \eqref{eq:problem:31} is still nonconvex due to its non-concave objective function and  nonconvex constraints \eqref{eq:problem:31:b}-\eqref{eq:problem:31:e}. To make the problem tractable, we introduce the additional convex constraints as
\begin{IEEEeqnarray}{rCl} \label{eq:peruser pergroup rate with time}
	&& t_g\hat{\tau}_k^g\geq \bigl(\tilde{\tau}_k^g\bigr)^2,\,\forall k\in\mathcal{K}, \forall g\in\mathcal{G},\IEEEyessubnumber \label{eq:peruser pergroup downlink rate with time} \\
	&& t_g\hat{\kappa}_\ell^g\geq \bigl(\tilde{\kappa}_\ell^g\bigr)^2,\,\forall \ell\in\mathcal{L}, \forall g\in\mathcal{G}, \IEEEyessubnumber\label{eq:peruer pergroup uplink with time}
\end{IEEEeqnarray}
where  $ \boldsymbol{\tilde{\tau}} \triangleq \bigl[\tilde{\tau}_k^g\bigr]_{k\in\mathcal{K},g\in\mathcal{G}} $ and $ \boldsymbol{\tilde{\kappa}} \triangleq \bigl[\tilde{\kappa}_\ell^g\bigr]_{\ell\in\mathcal{L},g\in\mathcal{G}} $  are newly introduced variables. Owing to the concavity of $\bigl(\tilde{\tau}_k^g\bigr)^2$ and $\bigl(\tilde{\kappa}_\ell^g\bigr)^2$, we can iteratively approximate them as the first-order approximation.  At the feasible point $(\tilde{\tau}_k^{g,(n)},\tilde{\kappa}_k^{g,(n)})$, $\bigl(\tilde{\tau}_k^g\bigr)^2$ and $\bigl(\tilde{\kappa}_\ell^g\bigr)^2$ are lower bounded by
\begin{IEEEeqnarray}{rCl} \label{eq:peruser pergroup rate with time approx.}
	 \bigl(\tilde{\tau}_k^g\bigr)^2 &\geq&
	\bigl(\tilde{\tau}_k^{g,(n)}\bigr)^2 + 2\tilde{\tau}_k^{g,(n)}\bigl(\tilde{\tau}_k^{g}-\tilde{\tau}_k^{g,(n)}\bigr) \nonumber\\
	 &:=& \mathcal{\widetilde{F}}_k^{g,(n)}(\tilde{\tau}_k^g),\IEEEyessubnumber \label{eq:peruser pergroup downlink rate with time approx.} \\
	 \bigl(\tilde{\kappa}_\ell^g\bigr)^2 &\geq& \bigl(\tilde{\kappa}_\ell^{g,(n)}\bigr)^2 + 2\tilde{\kappa}_\ell^{g,(n)}\bigl(\tilde{\kappa}_\ell^{g}-\tilde{\kappa}_\ell^{g,(n)}\bigr) \nonumber\\
	&:=& \mathcal{\widetilde{P}}_\ell^{g,(n)}(\tilde{\kappa}_\ell^g). \label{eq:peruser pergroup uplink with time approx.}\IEEEyessubnumber
\end{IEEEeqnarray}
With \eqref{eq:peruser pergroup rate with time approx.}, the objective functions in \eqref{eq:problem:31} are concave. Obviously, the constraints \eqref{eq:problem:31:b} and \eqref{eq:problem:31:c} are also convex.

We  are now in position to further expose the hidden convexity of the nonconvex constraints \eqref{eq:problem:31:d} and \eqref{eq:problem:31:e}. Since they are of the same type, let us treat \eqref{eq:problem:31:d} first by using the following chains:
\begin{IEEEeqnarray}{rCl}\label{eq:problem:31:d1}
	\sum_{k=1}^K\|\mathbf{w}_k^g\|^2 &\leq& \omega_g,\;\forall g \in\mathcal{G},\IEEEyessubnumber \label{eq:problem:31:d2} \\
	 \sum_{g=1}^Gt_g\omega_g &\leq& P_{bs},\IEEEyessubnumber \label{eq:problem:31:d3}
\end{IEEEeqnarray}
where $ \boldsymbol{\omega} \triangleq \bigl[\omega_g\bigr]_{g\in\mathcal{G}} $  are newly introduced variables. For the nonconvex constraint \eqref{eq:problem:31:d3},  a convex upper bound of   $\xi^{g}(t_g,\omega_g)\triangleq t_g\omega_g$ can be found as \cite{Dan:TWC:14}
\begin{IEEEeqnarray}{rCl}\label{eq:problem:31:d4}
	 t_g \omega_g &\leq& \frac{1}{2}\frac{\left(t_g\right)^2}{r^{(n)}(t_g,\omega_g)}+ \frac{1}{2}\left(\omega_g\right)^2 r^{(n)}(t_g,\omega_g) \nonumber\\
	&:=& \xi^{g,(n)}(t_g,\omega_g), 
\end{IEEEeqnarray}
where  $ r^{(n)}(x,y) $ is defined as $ r^{(n)}(x,y) \triangleq x^{(n)}/y^{(n)} $. It is readily seen that \eqref{eq:problem:31:d4} holds with equality at optimum. Thus, \eqref{eq:problem:31:d3} is transformed to the following convex constraint:
\begin{IEEEeqnarray}{rCl}\label{eq:problem:31:d5}
	 \sum_{g=1}^{G} \xi^{g,(n)}(t_g,\omega_g) \leq P_{bs}.
\end{IEEEeqnarray}
By introducing the new variables  $ \boldsymbol{\hat{p}} \triangleq \bigl[\hat{p}^g_{\ell}\bigr]_{\ell\in\mathcal{L},g\in\mathcal{G}}$, the last constraint \eqref{eq:problem:31:e} can be shaped to take the following constraints:
\begin{IEEEeqnarray}{rCl}\label{eq:problem:31:e1}
	(p^g_{\ell})^2 &\leq& \hat{p}^g_{\ell},\;\forall \ell \in\mathcal{L},\forall g \in\mathcal{G},\IEEEyessubnumber \label{eq:problem:31:e2} \\
	 \sum_{g=1}^Gt_g\hat{p}_{\ell}^g &\leq& P_{\ell},\,\forall \ell\in\mathcal{L},\IEEEyessubnumber \label{eq:problem:31:e3}
\end{IEEEeqnarray}
Invoking the same procedure to \eqref{eq:problem:31:d4}, the nonconvex constraints \eqref{eq:problem:31:e3} can be safely approximated as
\begin{IEEEeqnarray}{rCl}\label{eq:power constraints convex}
  \sum_{g=1}^{G} \xi^{g,(n)}(t_g,\hat{p}_\ell^g) \leq P_\ell,\;\forall \ell \in\mathcal{L}.
\end{IEEEeqnarray}

With the above results, we now address the nonconvex optimization problem \eqref{eq:problem:1} by successively solving the following convex
quadratic program:
\begin{IEEEeqnarray}{rCl}\label{eq:problem:5}
\underset{\substack{\mathbf{w},  \mathbf{p},\boldsymbol{\phi}, \boldsymbol{\theta}, \tilde{\boldsymbol{\theta}}, \boldsymbol{\alpha},  \boldsymbol{\beta},\boldsymbol{t},\\ \boldsymbol{\tau},\hat{\boldsymbol{\tau}},\boldsymbol{\kappa},\hat{\boldsymbol{\kappa}},\tilde{\boldsymbol{\tau}},\tilde{\boldsymbol{\kappa}},\boldsymbol{\omega},\hat{\boldsymbol{p}}}}{\mathrm{maximize}}\ &&
\sum_{k=1}^K\sum_{g=1}^{G}\mathcal{\widetilde{F}}_k^{g,(n)}(\tilde{\tau}_k^g) + \sum_{\ell =1}^{L}\sum_{g=1}^{G}\mathcal{\widetilde{P}}_\ell^{g,(n)}(\tilde{\kappa}_\ell^g)    \IEEEyessubnumber \label{eq:problem:5:a}\qquad\;\;\\
                          \st \;												
						&& \sum\nolimits_{g=1}^{G}\mathcal{\widetilde{F}}_k^{g,(n)}(\tilde{\tau}_k^g) \geq \bar{R}^{\mathsf{D}}_k,\; \forall k\in\mathcal{K}, \IEEEyessubnumber \label{eq:problem:5:b}\\		
						&&	\sum\nolimits_{g=1}^{G}\mathcal{\widetilde{P}}_\ell^{g,(n)}(\tilde{\kappa}_\ell^g)\geq \bar{R}^{\mathsf{U}}_\ell,\, \forall \ell \in\mathcal{L}, \IEEEyessubnumber \label{eq:problem:5:c}\\
						&&		\eqref{eq:modified:pd}, \eqref{eq:modified:g}, \eqref{eq:10a:common}, \eqref{eq:convexappro:b2}, \eqref{eq:convexappro:c2},\eqref{eq:problem:3:b}, \eqref{eq:problem:3:c},\nonumber\\
						&& \eqref{eq:peruser pergroup rate},\eqref{eq:peruser pergroup rate approx.},\eqref{eq:peruser pergroup rate with time},\eqref{eq:problem:31:d2},\eqref{eq:problem:31:d5},\eqref{eq:problem:31:e2},\eqref{eq:power constraints convex}.\IEEEyessubnumber \label{eq:problem:5:d}
\end{IEEEeqnarray}

The iterative algorithm  outlined in Algorithm~\ref{algo:proposed:2} obtains the solution of the  convex optimization problem \eqref{eq:problem:5} in its $(n+1)$-th iteration. In a  manner similar to Proposition~\ref{prop1}, we
can show that Algorithm~\ref{algo:proposed:2} yields a sequence of improved points of \eqref{eq:problem:1} due to  updating
the involved variables after each iteration, which converges to a KKT point. The convex problem \eqref{eq:problem:5} involves  $( N_{\mathrm{tx}}K + 6K + 7L+ 2)G$ scalar real decision variables and $(7KG+7LG+K+2L+2G+2)$  quadratic and linear constraints, so its computational complexity in each iteration of Algorithm~\ref{algo:proposed:2} is $\mathcal{O}\Bigl((7KG+7LG+K+2L+2G+2)^{2.5}\bigl(( N_{\mathrm{tx}}K + 6K + 7L+ 2)^2G^2 + 7KG+7LG+K+2L+2G+2\bigr) \Bigr)$ \cite{Ben:2001}.

\begin{algorithm}[t]
\begin{algorithmic}[1]

\protect\caption{Proposed iterative algorithm for the joint SR maximization problem \eqref{eq:problem:1}}
\label{algo:proposed:2}

\global\long\def\algorithmicrequire{\textbf{Initialization:}}

\REQUIRE  Set $n:=0$, $\alpha_{k}^{g,(0)}=\beta_{\ell}^{g,(0)}=1/2$,  $t_g^{(0)}=1/G$, and solve \eqref{eq:convexappro:obj32} to generate an initial feasible point $(\mathbf{w}^{(0)},\mathbf{p}^{(0)},\boldsymbol{\phi}^{(0)},\boldsymbol{\tau}^{(0)},\boldsymbol{\kappa}^{(0)},\tilde{\boldsymbol{\tau}}^{(0)},\tilde{\boldsymbol{\kappa}}^{(0)},\boldsymbol{\omega}^{(0)},\hat{\boldsymbol{p}}^{(0)})$.

\REPEAT
\STATE Solve \eqref{eq:problem:5} to obtain the optimal solutions ($\mathbf{w}^*,  \mathbf{p}^*,\boldsymbol{\phi}^*, \boldsymbol{\theta}^*, \tilde{\boldsymbol{\theta}}^*, \boldsymbol{\alpha}^*,  \boldsymbol{\beta}^*,\boldsymbol{t}^*,\boldsymbol{\tau}^*,\hat{\boldsymbol{\tau}}^*,\boldsymbol{\kappa}^*,$ $\hat{\boldsymbol{\kappa}}^*,\tilde{\boldsymbol{\tau}}^*,\tilde{\boldsymbol{\kappa}}^*,\boldsymbol{\omega}^*,\hat{\boldsymbol{p}}^*$).

\STATE Update $\mathbf{w}^{(n+1)}:=\mathbf{w}^{*},$   $\mathbf{p}^{(n+1)}:=\mathbf{p}^{*}$,
   $ \boldsymbol{\phi}^{(n+1)}:=\boldsymbol{\phi}^{*}$, $ \boldsymbol{\tau}^{(n+1)}:=\boldsymbol{\tau}^{*}$,  $ \boldsymbol{\kappa}^{(n+1)}:=\boldsymbol{\kappa}^{*}$,  $ \tilde{\boldsymbol{\tau}}^{(n+1)}:=\tilde{\boldsymbol{\tau}}^{*}$,  $ \tilde{\boldsymbol{\kappa}}^{(n+1)}:=\tilde{\boldsymbol{\kappa}}^{*}$, $ \boldsymbol{\omega}^{(n+1)}:=\boldsymbol{\omega}^{*}$,  and $ \hat{\boldsymbol{p}}^{(n+1)}:=\hat{\boldsymbol{p}}^{*}$.
\STATE Set $n:=n+1.$
\UNTIL Convergence\\
\end{algorithmic} \end{algorithm}

\textit{Generation of initial points}: Initialized by any feasible $(\mathbf{w}^{(0)},\mathbf{p}^{(0)},\boldsymbol{\phi}^{(0)},\boldsymbol{\tau}^{(0)},\boldsymbol{\kappa}^{(0)},\tilde{\boldsymbol{\tau}}^{(0)},\tilde{\boldsymbol{\kappa}}^{(0)},$ $\boldsymbol{\omega}^{(0)},\hat{\boldsymbol{p}}^{(0)})$, we successively solve 
\begin{IEEEeqnarray}{rCl}\label{eq:convexappro:obj32}
 \ds\underset{\substack{\mathbf{w},  \mathbf{p},\boldsymbol{\phi}, \boldsymbol{\theta}, \tilde{\boldsymbol{\theta}}, \boldsymbol{\alpha},  \boldsymbol{\beta},\boldsymbol{t},\\ \boldsymbol{\tau},\hat{\boldsymbol{\tau}},\boldsymbol{\kappa},\hat{\boldsymbol{\kappa}},\tilde{\boldsymbol{\tau}},\tilde{\boldsymbol{\kappa}},\boldsymbol{\omega},\hat{\boldsymbol{p}}}}{\mathrm{maximize}}\; && \underset{\substack{k\in\mathcal{K}\\ \ell\in\mathcal{L}}}{\min}\;\Biggl\{\frac{\sum_{g=1}^{G}\mathcal{\widetilde{F}}_k^{g,(n)}(\tilde{\tau}_k^g)}{\bar{R}^{\mathsf{D}}_k},\nonumber\\
&&\qquad\quad\frac{\sum_{g=1}^{G}\mathcal{\widetilde{P}}_\ell^{g,(n)}(\tilde{\kappa}_\ell^g)}{\bar{R}^{\mathsf{U}}_\ell}    \Biggl\}\IEEEyessubnumber \label{eq:convexappro:a32}\\
\st && \quad  \eqref{eq:problem:5:d}, \IEEEyessubnumber \label{eq:convexappro:d32}
\end{IEEEeqnarray}
until reaching its optimal value of more than or equal to $1$ to satisfy the constraints in \eqref{eq:problem:1}, i.e.,
\begin{equation}\nonumber
\underset{\substack{k\in\mathcal{K}\\ \ell\in\mathcal{L}}}{\min}\;\Biggl\{\frac{\sum_{g=1}^{G}\mathcal{\widetilde{F}}_k^{g,(n)}(\tilde{\tau}_k^g)}{\bar{R}^{\mathsf{D}}_k},\frac{\sum_{g=1}^{G}\mathcal{\widetilde{P}}_\ell^{g,(n)}(\tilde{\kappa}_\ell^g)}{\bar{R}^{\mathsf{U}}_\ell}   \Biggl\} \geq 1.
\end{equation}

\begin{remark}
The optimization problem \eqref{eq:problem:5} requires additional number of variables, which may lead to a high computational complexity. However, Algorithm~\ref{algo:proposed:2} requires solving only a simple convex quadratic program for each iteration. In addition, all constraints in \eqref{eq:problem:5} are   linear or second-order cone (SOC) representable \cite[Sec. 3.3]{Ben:2001}. Thus, we are able to arrive at a SOC program (SOCP). This helps  reduce the overall run-time for Algorithm~\ref{algo:proposed:2}
due  to extremely efficient state-of-the-art SOCP solvers.
\end{remark}

\subsection{Additional Conditions for Grouping Assignment}

As mentioned earlier, the value of $ \boldsymbol{\alpha} $ ($ \boldsymbol{\beta} $) indicates that the DLUs (ULUs) are active or inactive at a specified group. However, we have numerically observed that there exists a case where the downlink or uplink power for a user at a certain group is nearly zero, which is associated with an inactive user. Thus, the optimal solution can be obtained no matter how $ \alpha_k^g $ or $ \beta_\ell^g $ assigned to that user has been chosen. To manage the assignments exactly, we put the following additional linear constraints:
\begin{IEEEeqnarray}{rCl}\label{eq:assigningconstraint}
	&& \alpha_k^g \leq \Omega \ddot{\mathcal{F}}_k^{g,(n)}(\mathbf{w}, \boldsymbol{\theta}), \IEEEyessubnumber \label{eq:assigningconstraint:alpha}\\
	&& \beta_\ell^g \leq \Omega \ddot{\mathcal{P}}_\ell^{g,(n)}(\mathbf{p},\tilde{\boldsymbol{\theta}}), \IEEEyessubnumber \label{eq:assigningconstraint:beta}
\end{IEEEeqnarray}
where $ \Omega $  is a given constant. The value of $ \Omega $ needs to be large enough to force $ \alpha_k^g $ and $ \beta_\ell^g $ to reach 0 or 1 quickly. In fact, when the rate of a user in a group is vital to the sum rate, the grouping assignment variables satisfy
\begin{IEEEeqnarray}{rCl}
  0 \leq \alpha_k^g &\leq& 1 \leq \Omega \ddot{\mathcal{F}}_k^{g,(n)}(\mathbf{w}, \boldsymbol{\theta}), \IEEEyessubnumber \\
  0 \leq \beta_\ell^g &\leq& 1 \leq \Omega \ddot{\mathcal{P}}_\ell^{g,(n)}(\mathbf{p},\tilde{\boldsymbol{\theta}}),\IEEEyessubnumber   
\end{IEEEeqnarray}
to rapidly boost $ \alpha_k^g $ and $ \beta_\ell^g $ up to 1. Otherwise, they are depressed to 0 by warranting 
\begin{IEEEeqnarray}{rCl}
 0 \leq \alpha_k^g &\leq& \Omega \ddot{\mathcal{F}}_k^{g,(n)}(\mathbf{w}, \boldsymbol{\theta}) \leq 1, \IEEEyessubnumber \\
 0 \leq \beta_\ell^g &\leq& \Omega \ddot{\mathcal{P}}_\ell^{g,(n)}(\mathbf{p},\tilde{\boldsymbol{\theta}}) \leq 1, \IEEEyessubnumber 
\end{IEEEeqnarray}
when $ \ddot{\mathcal{F}}_k^{g,(n)}(\mathbf{w}, \boldsymbol{\theta}) $ and $\ddot{\mathcal{P}}_\ell^{g,(n)}(\mathbf{p},\tilde{\boldsymbol{\theta}}) $ are negligible.

\section{Numerical Results}\label{NumericalResults}
 
\begin{table}[t]
\caption{Simulation Parameters}
	\label{parameter}
	\centering
		\begin{tabular}{l|l}
		\hline
				Parameter & Value \\
		\hline\hline
		    Carrier frequency/ Bandwidth                            & 2 GHz/ 10 MHz \\
				Radius of small cell ($r$)                    & 100 m \\
				Distance between the BS and  nearest user  & $\geq$ 10 m\\
				Noise power spectral density $(\sigma^2_{k}, \sigma^2)$& -174 dBm/Hz \\
				Path loss from the  BS to a user  $(\sigma_{\mathsf{LOS}})$ & 103.8 + 20.9$\log_{10}(d)$ dB\\
				Path loss from the $\mathsf{U}_{\ell}$ to the $\mathsf{D}_k$	$(\tilde{\sigma}_{\mathsf{NLOS}})$& 145.4 + 37.5$\log_{10}(\hat{d})$ dB\\
				Power budge at the BS ($P_{bs}$)     & 26 dBm, as in \cite{3GPP} \\
				Power budge at  $\mathsf{U}_\ell$ ($P_{\ell},\forall \ell$)     & 10 dBm, as in \cite{Duarte:TWC:12}  \\
			  FD residual SI ($\rho$) & -75 dBm, as in \cite{Bharadia13} \\
			  Predetermined rate threshold ($\bar{\mathsf{R}}$) & 1 bps/Hz\\
				Number of antennas at  BS ($N_{\mathrm{tx}} = N_{\mathrm{rx}})$ & 4\\
		\hline		   				
		\end{tabular}
\end{table}

We now evaluate the numerical performance of the proposed algorithms using  computer simulations.
The channel vectors from the BS to a DLU,  from a ULU to the BS, and from the ULU to DLU  are assumed to undergo the path loss model for line-of-sight (LOS) and non-line-of-sight (NLOS) scenarios, respectively  \cite{Tam:TCOM:16,Dan:TWC:14,3GPP}.
Specifically, the channel vector from the BS to $\mathsf{D}_k$ is modeled as $\mathbf{h}_{k}=\sqrt{\mathsf{PL}_{\mathsf{D}_k}}\tilde{\mathbf{h}}_{k}$, where the entries of $\tilde{\mathbf{h}}_{k}$ are generated as independent circularly symmetric complex Gaussian (CSCG) random variables with distribution $\mathcal{CN}(0,1)$, and $\mathsf{PL}_{\mathsf{D}_k}=10^{(-\sigma_{\mathsf{LOS}}/10)}$ represents the path loss. The channel vectors from  $\mathsf{U}_{\ell}$ to the BS and $\mathsf{U}_{\ell}$ to $\mathsf{D}_{k}$ are generated similarly as $\mathbf{g}_{\ell}=\sqrt{\mathsf{PL}_{\mathsf{U}_{\ell}}}\tilde{\mathbf{g}}_{\ell}$ and  $\hat{g}_{\ell k}=\sqrt{\mathsf{PL}_{\ell k}}\tilde{g}_{\ell k}$, where the entries of $\tilde{\mathbf{g}}_{\mathsf{U}_{\ell}}$ and $\tilde{g}_{\ell k}$ are  independent CSCG random variables with distribution $\mathcal{CN}(0,1)$, and their  path losses are $\mathsf{PL}_{\mathsf{U}_{\ell}} = 10^{(-\sigma_{\mathsf{LOS}}/10)}$ and $\mathsf{PL}_{\ell k} = 10^{(-\sigma_{\mathsf{NLOS}}/10)}$. The entries of the fading loop channel $\mathbf{G}_{\mathsf{I}}$ are independently  drawn from the CSCG distribution $\mathcal{CN}(0, 1)$ \cite{Riihonen-SP-11}. Unless stated otherwise, the  parameters are given in Table~\ref{parameter}  for ease of cross referencing.  In Table \ref{parameter}, $d$ ($\hat{d}$)
is the distance between the BS and a user (between the ULU and DLU).  We simulate small-cell scenarios where all users are randomly placed in a circle area of a radius $r = 100$ m. Without loss of generality, we set the  predetermined rate threshold of
all users to $\bar{\mathsf{R}} = \bar{R}^{\mathsf{D}}_k = \bar{R}^{\mathsf{U}}_\ell$ and all  ULUs are assume to have the same maximum transmit power.  The convex solver that is used is SDPT3 \cite{Toh} with  the parser YALMIP \cite{Lofberg} in the MATLAB environment. The error
tolerances of all iterative algorithms are set to $\epsilon_{\mathsf{err}}  = 10^{-3}$. We divide the achieved SR results by $\ln(2)$ to arrive at the unit of bps/channel-use in binary communications.

\begin{figure}[t]
\centering
        \includegraphics[width=0.49\textwidth]{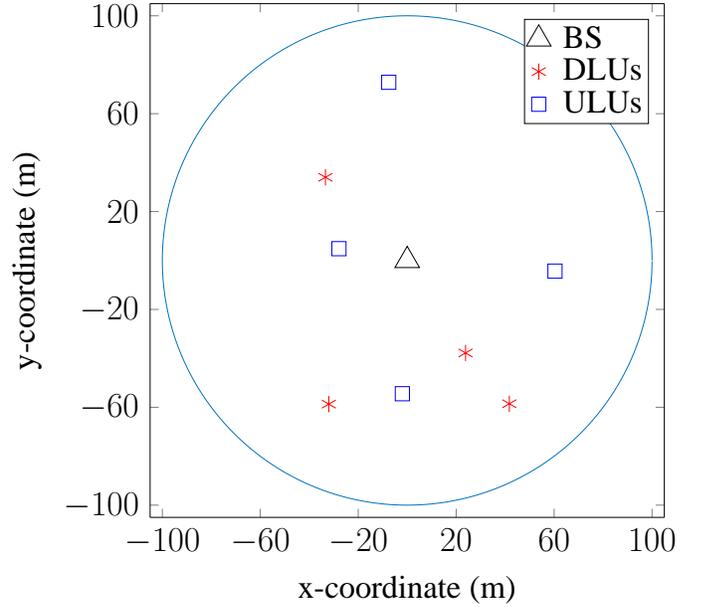}
	  \caption{Location of users for the simulation setup used in Figs.~\ref{fig:Convergencebehavior}-\ref{fig:NoUE}. The users are assumed to be uniformly distributed over the cell area.}\label{fig:celllayout}
\end{figure}

\begin{figure}[t]
    	\centering
        \includegraphics[width=0.48\textwidth]{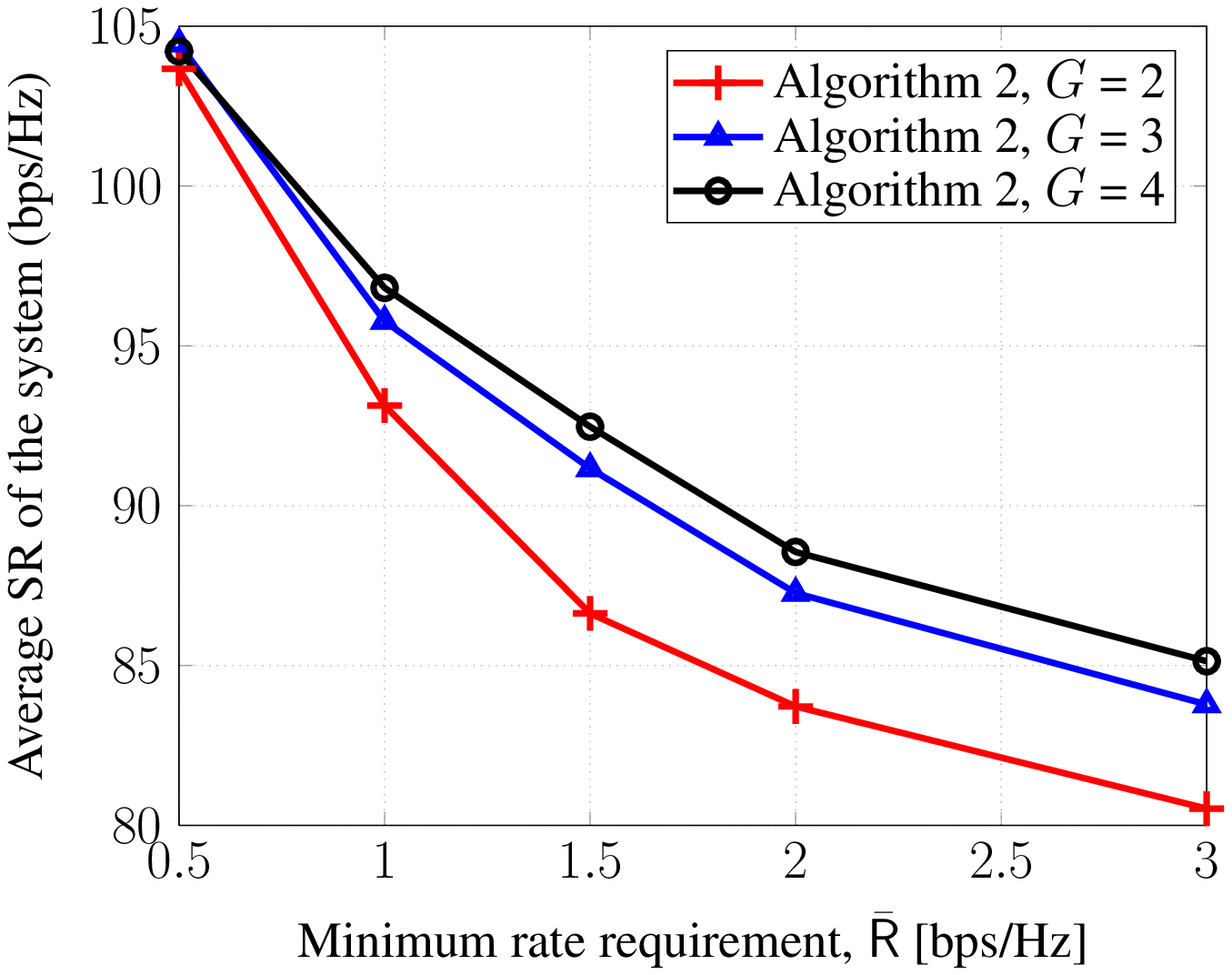}
        \caption{Average SR versus  $\bar{\mathsf{R}}$  with $N_{\mathrm{tx}} = N_{\mathrm{rx}}$ = 10.}\label{fig:SRvsR10UE}
\end{figure}

We also compare the performance of our proposed FD system
with those of the HD system and FD system \cite{Dan:TWC:14}. For the HD system,  the BS uses all  antennas, i.e., $N_{\mathrm{tx}} + N_{\mathrm{rx}}$, for communication in each direction. Suppose the achieved SRs of the uplink and downlink transmissions are $R_{\mathsf{UL}}$ and $R_{\mathsf{DL}}$, respectively, which are computed independently. Consequently, the total SR of the HD system per resource block is calculated as $R_{\mathsf{HD}} = (R_{\mathsf{UL}}+R_{\mathsf{DL}})/2$. For these cases, the proposed Algorithm~\ref{algo:proposed:DUAL} also provides the optimal solution for both the HD system and FD system  \cite{Dan:TWC:14} by simply setting $G = 1$ (no user grouping). The numbers of transmit and receive antennas at the BS are set to $N_{\mathrm{tx}} = N_{\mathrm{rx}} = 4$, except for Fig.~\ref{fig:SRvsR10UE}, in which $N_{\mathrm{tx}} = N_{\mathrm{rx}} = 10$ are used. The simulation results are derived by averaging over 100 runs for different locations of users, as depicted in Fig.~\ref{fig:celllayout}.

\begin{figure}[t]
\centering
        \includegraphics[width=0.49\textwidth]{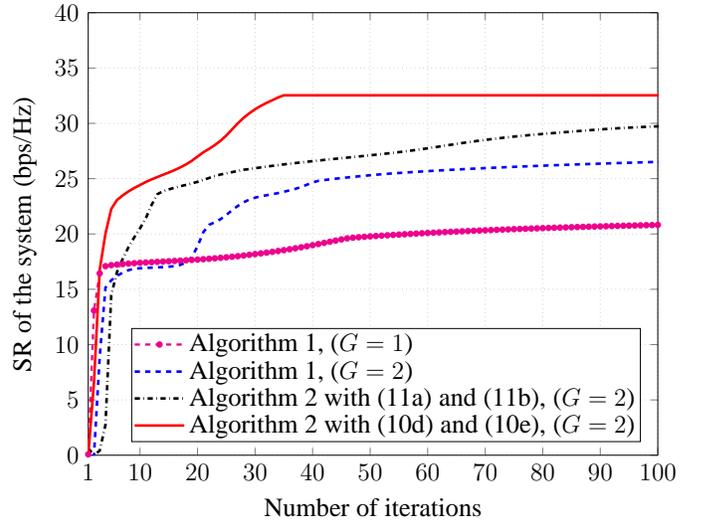}
	  \caption{Typical convergence behavior  of Algorithm~\ref{algo:proposed:DUAL} and Algorithm~\ref{algo:proposed:2}.}\label{fig:Convergencebehavior}
\end{figure}

Fig.~\ref{fig:Convergencebehavior}
illustrates the typical convergence behavior of the proposed Algorithms~\ref{algo:proposed:DUAL} and \ref{algo:proposed:2} for the number of groups and different types of power constraints. Both algorithms converge within tens of iterations for all cases. We can see that as increase in the number of groups results in a higher  system performance in terms of the SR. In addition, the proposed Algorithm~\ref{algo:proposed:2} with real power constraints (i.e., \eqref{eq:modified:b} and \eqref{eq:modified:c}) and additional constraints (i.e., \eqref{eq:assigningconstraint:alpha} and \eqref{eq:assigningconstraint:beta}) requires 35 iterations, which is the fastest convergence rate when compared to others.
Of course, Algorithm~\ref{algo:proposed:2} using the real power constraints \eqref{eq:modified:b} and \eqref{eq:modified:c} offers better SR compared to that of using the relaxed power constraints \eqref{eq:PT:conv:BSa} and \eqref{eq:PT:conv:BSb}. Note that the left-hand side (LHS) of \eqref{eq:PT:conv:BS}  is a sum of power consumptions only, while the LHS of \eqref{eq:modified:b} and \eqref{eq:modified:c}  is the total transmit power at the BS. Consequently, by using \eqref{eq:PT:conv:BS}, the BS and ULUs do not use all allowable power. Thus, the corresponding performance is not optimal. In the following simulation results, we use the power constraints \eqref{eq:modified:b} and \eqref{eq:modified:c} instead of using \eqref{eq:PT:conv:BSa} and \eqref{eq:PT:conv:BSb}.

\begin{figure}[t]
    	\centering
        \includegraphics[width=0.49\textwidth]{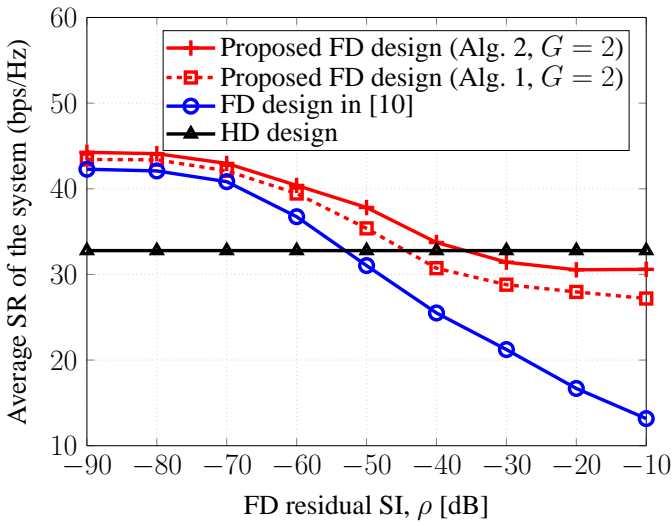}
        \caption{Average SR versus $\rho$  with $K = L = 4$.}\label{fig:SRvsSI}
\end{figure}

In Fig.~\ref{fig:SRvsSI}, we depict  the effect of the FD residual SI.  When the SI is sufficiently canceled $(\rho \leq -75$ dB), the SRs of FD systems are  better than those   of the HD counterpart by about $34.50\%$, $32,34\%$, and $28.37\%$, corresponding to the proposed FD system with Algorithm~\ref{algo:proposed:2}, Algorithm~\ref{algo:proposed:DUAL}, and the FD system \cite{Dan:TWC:14}, respectively.  Generally, the SR of the FD systems  is degraded as the residual SI becomes larger $(\rho > -70$ dB) and tends to be worse than the HD system for $\rho > -35$ dB. This is probably attributed to the fact that the BS needs to scale down the transmit power in the downlink transmission to avoid a harmful effect on the uplink channel, which results in the SR loss of the
FD system. As expected, the FD design with joint optimization of the fraction of time and user grouping assignment in Algorithm~\ref{algo:proposed:2} offers better  SR compared to the fixed design in Algorithm~\ref{algo:proposed:DUAL} as a result of the optimized transmission. Another interesting observation is that for whatever SI level, the SR of the proposed FD system still outperforms the  FD system in \cite{Dan:TWC:14}, and its gain increases  when $\rho$ becomes large. 
 The simulation results in Fig.~\ref{fig:SRvsSI} further confirm that the proposed user grouping allows us to exploit the multiuser diversity gain in both channels and also mitigate the effect of the FD residual SI. Due to the advantages of Algorithm~\ref{algo:proposed:2}, from now on, we  plot the system performance only with the algorithm.

\begin{figure}[t]
    	\centering
        \includegraphics[width=0.49\textwidth]{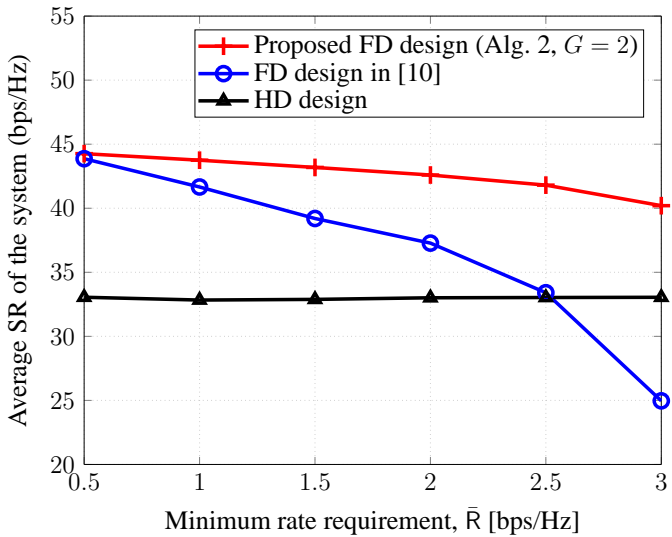}
        \caption{Average SR versus  $\bar{\mathsf{R}}$  with $K = L = 4$.}\label{fig:SRvsR}
\end{figure}

The  SR  versus the minimum rate requirement $\bar{\mathsf{R}}\in [0.5,\ 3]$ bps/Hz for all users   is shown in Fig.~\ref{fig:SRvsR}. This shows that the HD system is fulfilled while the SR of FD design in \cite{Dan:TWC:14} is degraded dramatically when $\bar{\mathsf{R}}$ increases. Notably, the proposed FD design is slightly degraded when $\bar{\mathsf{R}}$ becomes higher. In fact, the SR of the system is mostly a contribution of the downlink channel, and thus, the BS must pay more attention to serving ULUs when $\bar{\mathsf{R}}$ increases by reducing its transmit power.
Recalling the discussion from Fig.~\ref{fig:SRvsSI}, the proposed FD design exploits the multiuser diversity gain more efficiently, which in turn improves the user fairness, as shown in Fig.~\ref{fig:SRvsR}. Certainly, the proposed FD system  achieves better
SR than that of the FD in \cite{Dan:TWC:14}, and the gap between the two is even deeper.

\begin{figure}[t]
    	\centering
        \includegraphics[width=0.49\textwidth]{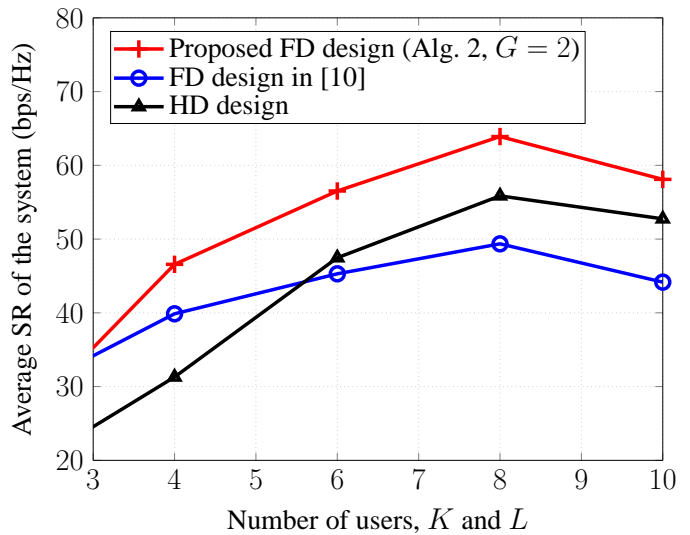}
        \caption{Average SR versus the number of users $K$ and $L$ with $K=L$.}\label{fig:NoUE}
\end{figure}

We plot the impact of the number of  users on the system performance  in Fig.~\ref{fig:NoUE}. A multiuser system is considered to demonstrate that the proposed FD system can deliver higher SR than conventional FD and HD systems  for $K=L \in[3,\,10]$.  The SRs of all designs first increase as $K$ and $L$ increase up to 8 for a given $N_{\mathrm{tx}} = N_{\mathrm{rx}} = 4$ and then decrease  with $K$ and $L$ because  it lacks the degree-of-freedom (DoF) for leveraging multiuser diversity. 
Of course, the optimal value of $K$ and $L$  may be different in other settings. For $K \geq 6$ and $L \geq 6$,  the HD system achieves a better SR than the FD system in \cite{Dan:TWC:14} because of the fact that the BS uses all available antennas ($N_{\mathrm{tx}} + N_{\mathrm{rx}}$) leading to more DoFs available for transmission. Again,  the SR of the proposed FD system outperforms that of the others. Thus,  the proposed user grouping is a powerful means to combating the DoF bottleneck. 

\begin{figure}[t]
\centering
        \includegraphics[width=0.475\textwidth]{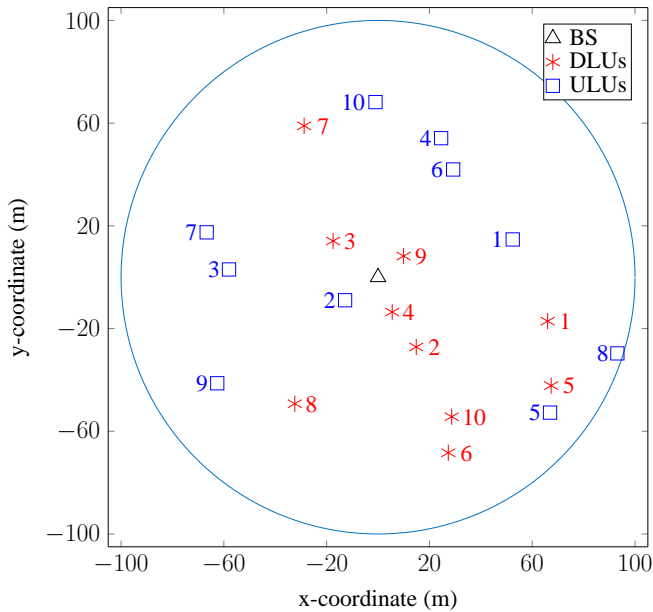}
	  \caption{Location of users for the simulation setup used in Fig.~\ref{fig:SRvsR10UE} and Table~\ref{lab:ratespergroups}. The numbers of DLUs and ULUs are set to $K = L = 10$.}\label{fig:celllayoutTableII}
\end{figure}

In Fig.~\ref{fig:SRvsR10UE}, we investigate the impact of the number of groups $G$ on the system performance for the simulation setup given in Fig.~\ref{fig:celllayoutTableII}. We plot the SR of the proposed FD system versus the minimum rate requirement $\bar{\mathsf{R}}\in [0.5,\ 3]$ bps/Hz for  $N_{\mathrm{tx}} = N_{\mathrm{rx}}$ = 10. As expected, the SR increases as the number of groups increases, although the gain tends to diminish with $G$. This implies that the larger number of groups would be appropriate for the larger number of users.

Finally, we provide further insight on the proposed FD system by presenting how the users are grouped for the simulation setup given in Fig.~\ref{fig:celllayoutTableII}. There are 20 users in total with 10 DLUs and 10 ULUs for $\bar{\mathsf{R}} = 0.5$ bps/Hz and $N_{\mathrm{tx}} = N_{\mathrm{rx}}$ = 4. The resulting rates in Table~\ref{lab:ratespergroups}  show that the users are divided into different groups. Looking at the $\mathsf{U}_{5}$, for instance, we can see that its location is very close to the DLUs $\{\mathsf{D}_{1}$, $\mathsf{D}_{5}$, $\mathsf{D}_{6}$  $\mathsf{D}_{10}\}$ in Fig.~\ref{fig:celllayoutTableII}. Therefore, $\mathsf{U}_{5}$ may cause a large amount of interference to those DLUs. This is the reason why $\mathsf{U}_{5}$ belongs to a group different from such near DLUs, as manifested in Table~\ref{lab:ratespergroups}. Intuitively, DLUs are served in only one specific group due to reducing the CCI and  of course they  can be changed in other settings. On the other hand, the FD system in \cite{Dan:TWC:14} favors the users \{$\mathsf{U}_{1}$, $\mathsf{U}_{2}$, $\mathsf{U}_{3}$, $\mathsf{U}_{9}$\} in good channel conditions   while barely meeting the rate requirements of the other users. The proposed algorithm is shown to provide better user fairness. Moreover, the proposed FD system yields the SR that is  11.29 bps/Hz higher than the FD system in \cite{Dan:TWC:14}.

\begin{table}[t]
\centering
\caption{Achieved User Rates (bps/Hz) in Different Groups and Comparison with FD in \cite{Dan:TWC:14} for $\bar{\mathsf{R}} = 0.5$ bps/Hz}
\label{lab:ratespergroups}
{\setlength{\tabcolsep}{0.695em}
\setlength{\extrarowheight}{0.65em}
\begin{tabular}{|c|c|c|c|c|c|c| }
\hline
\multicolumn{2}{|c|}{\multirow{2}{*}{Users}}  & \multicolumn{4}{c|}{Algorithm 2 ($G = 3$)}                                                                                                                                                                                & FD in \cite{Dan:TWC:14} \\ \cline{3-7} 
\multicolumn{2}{|c|}{}                            & \begin{tabular}[c]{@{}l@{}}Group-1 \\ \end{tabular} & \begin{tabular}[c]{@{}l@{}}Group-2\\  \end{tabular} & \begin{tabular}[c]{@{}l@{}}Group-3\\ \end{tabular} & $\substack{\ds\text{Per-user}\\ \ds\text{Rate}}$ & $\substack{\ds\text{Per-user}\\ \ds\text{Rate}}$  \\ \hline \hline	
\multirow{10}{*}{DLUs}                       & 1  & 0.500                                                             & 0.000                                                             & 0.000                                                             & 0.500         & 0.500          \\ \cline{2-7} 
                                             & 2  & 0.000                                                             & 2.283                                                             & 0.000                                                             & 2.283         & 0.500          \\ \cline{2-7} 
                                             & 3  & 0.000                                                             & 4.042                                                             & 0.000                                                             & 4.042         & 0.500          \\ \cline{2-7} 
                                             & 4  & 0.000                                                             & 0.000                                                             & 1.717                                                             & 1.717         & 0.500          \\ \cline{2-7} 
                                             & 5  & 0.000                                                             & 0.000                                                             & 2.029                                                             & 2.029         & 0.500          \\ \cline{2-7} 
                                             & 6  & 0.687                                                             & 0.000                                                             & 0.000                                                             & 0.687         & 0.500          \\ \cline{2-7} 
                                             & 7  & 0.000                                                             & 0.501                                                             & 0.000                                                             & 0.501         & 0.500          \\ \cline{2-7} 
                                             & 8  & 0.544                                                             & 0.000                                                             & 0.000                                                             & 0.544         & 0.500          \\ \cline{2-7} 
                                             & 9  & 0.000                                                             & 0.000                                                             & 3.154                                                             & 3.154         & 0.500          \\ \cline{2-7} 
                                             & 10 & 1.283                                                             & 0.000                                                             & 0.000                                                             & 1.283         & 0.500          \\ \hline
					 \hline																	
\multicolumn{1}{|c|}{\multirow{10}{*}{ULUs}} & 1  & 0.695                                                             & 2.407                                                             & 0.570                                                             & 3.672         & 5.116          \\ \cline{2-7} 
\multicolumn{1}{|c|}{}                       & 2  & 2.114                                                             & 5.118                                                             & 3.543                                                             & 10.775        & 11.422         \\ \cline{2-7} 
\multicolumn{1}{|c|}{}                       & 3  & 0.668                                                             & 0.714                                                             & 1.295                                                             & 2.677         & 4.490          \\ \cline{2-7} 
\multicolumn{1}{|c|}{}                       & 4  & 0.438                                                             & 0.102                                                             & 0.813                                                             & 1.353         & 0.500          \\ \cline{2-7} 
\multicolumn{1}{|c|}{}                       & 5  & 0.000                                                             & 0.500                                                             & 0.000                                                             & 0.500         & 0.500          \\ \cline{2-7} 
\multicolumn{1}{|c|}{}                       & 6  & 0.000                                                             & 0.248                                                             & 0.252                                                             & 0.500         & 0.500          \\ \cline{2-7} 
\multicolumn{1}{|c|}{}                       & 7  & 0.173                                                             & 0.094                                                            & 1.184                                                             & 1.451         & 0.500          \\ \cline{2-7} 
\multicolumn{1}{|c|}{}                       & 8  & 0.000                                                             & 1.342                                                             & 0.089                                                             & 1.432         & 0.500          \\ \cline{2-7} 
\multicolumn{1}{|c|}{}                       & 9  & 1.083                                                             & 2.899                                                             & 2.197                                                             & 6.179         & 6.201          \\ \cline{2-7} 
\multicolumn{1}{|c|}{}                       & 10 & 0.776                                                             & 1.408                                                             & 1.147                                                             & 3.331         & 2.594          \\ \hline
            \hline	
\multicolumn{5}{|c|}{Total Sum Rate}                                                                                                                                                                                                                          & $\boldsymbol{\underline{48.612}}$        & $\boldsymbol{\underline{37.322}}$
         \\ \hline
\end{tabular}}
\end{table}

\section{Conclusion}\label{Conclusion}
In this paper, we have studied an FD system with joint user grouping,  time allocation, beamforming design, and power allocation optimization. To solve the original nonconvex optimization problem, we proposed new iterative algorithms to maximize the total sum rate of the system.
The proposed optimization problem captures all important factors in the system performance into low-complexity algorithms.
Numerical results with realistic parameters have confirmed that the proposed algorithms are guaranteed to  converge  to at least the local optima of the original nonconvex design problems. These results have been presented to show a fast convergence rate and to demonstrate the advantages of our proposed algorithms. The sum rate of the proposed FD system has been shown to be remarkably lager than HD if the residual SI is effectively canceled and to always outperform the FD system in \cite{Dan:TWC:14} with no user grouping in all cases. In addition, the proposed user grouping allows the FD system to exploit the multiuser diversity gain in both directions more efficiently and also to mitigate the effect of the FD residual SI and the CCI.

\section*{Appendix: Proof of Proposition \ref{prop1}}
 Let  $\mathcal{R}(\mathbf{w},\mathbf{p})$ and $\mathcal{R}^{(n)}(\mathbf{w},\mathbf{p},\boldsymbol{\phi})$ denote the objective values of \eqref{eq:problem:2} and \eqref{eq:convexappro:obj2}, respectively.
We have
\begin{equation}
\mathcal{R}(\mathbf{w},\mathbf{p})\geq
\mathcal{R}^{(n)}(\mathbf{w},\mathbf{p},\boldsymbol{\phi})\ \ \text{due to \eqref{eq:10a:3a} and \eqref{eq:10a:7a}},
\end{equation}
and
\begin{IEEEeqnarray}{rCl}
\mathcal{R}(\mathbf{w}^{(n)},\mathbf{p}^{(n)})&=&
\mathcal{R}^{(n)}(\mathbf{w}^{(n)},\mathbf{p}^{(n)},\boldsymbol{\phi}^{(n)})\ \nonumber\\
&&\text{due to \eqref{eq:10a:5} and \eqref{eq:10a:9}} .
\end{IEEEeqnarray}
The optimal solutions that are readily seen to return at the $n$-th iteration are also feasible for the considered problem at the $(n+1)$-th iteration. Let $\bigr(\mathbf{w}^{(n+1)},\mathbf{p}^{(n+1)},\boldsymbol{\phi}^{(n+1)}\bigr)$ and $\bigr(\mathbf{w}^{(n)},\mathbf{p}^{(n)},\boldsymbol{\phi}^{(n)}\bigr)$ be the optimal solutions of \eqref{eq:convexappro:obj2} at the $(n+1)$-th and at the $n$-th iteration, respectively. It follows that
\begin{IEEEeqnarray}{rCl}
\mathcal{R}(\mathbf{w}^{(n+1)},\mathbf{p}^{(n+1)})&\geq& \mathcal{R}^{(n)}(\mathbf{w}^{(n+1)},\mathbf{p}^{(n+1)},\boldsymbol{\phi}^{(n+1)})\nonumber\\
&\geq&\mathcal{R}^{(n)}(\mathbf{w}^{(n)},\mathbf{p}^{(n)},\boldsymbol{\phi}^{(n)})\nonumber\\
&=&\mathcal{R}(\mathbf{w}^{(n)},\mathbf{p}^{(n)}).\label{eq:appA}
\end{IEEEeqnarray}
 The inequalities in \eqref{eq:appA}  show that $\bigr(\mathbf{w}^{(n+1)},\mathbf{p}^{(n+1)},\boldsymbol{\phi}^{(n+1)}\bigr)$ is an improved point to \eqref{eq:convexappro:obj2} rather than $\bigr(\mathbf{w}^{(n)},\mathbf{p}^{(n)},\boldsymbol{\phi}^{(n)}\bigr)$ in the sense of increasing  objective value. In addition, the sequence of the objective is bounded above due to the power constraints in \eqref{eq:problem:b} and \eqref{eq:problem:c}. By following the same arguments as those in \cite[Theorem 1]{Marks:78}, we can prove that Algorithm~\ref{algo:proposed:DUAL} converges to a KKT point of \eqref{eq:problem:2}. 
Furthermore,  Algorithm~\ref{algo:proposed:DUAL} will terminate after a finite number of iterations, when it satisfies
\[\left| \frac{\mathcal{R}(\mathbf{w}^{(n+1)},\mathbf{p}^{(n+1)})-\mathcal{R}(\mathbf{w}^{(n)},\mathbf{p}^{(n)})}{\mathcal{R}(\mathbf{w}^{(n)},\mathbf{p}^{(n)})}\right| \leq \epsilon_{\mathsf{err}},
\]
where $\epsilon_{\mathsf{err}} > 0$ is a given tolerance.  Proposition \ref{prop1} is thus proved.

\balance
\bibliographystyle{IEEEtran}
\bibliography{IEEEFull}

\begin{thebibliography}{10}
\providecommand{\url}[1]{#1}
\csname url@samestyle\endcsname
\providecommand{\newblock}{\relax}
\providecommand{\bibinfo}[2]{#2}
\providecommand{\BIBentrySTDinterwordspacing}{\spaceskip=0pt\relax}
\providecommand{\BIBentryALTinterwordstretchfactor}{4}
\providecommand{\BIBentryALTinterwordspacing}{\spaceskip=\fontdimen2\font plus
\BIBentryALTinterwordstretchfactor\fontdimen3\font minus
  \fontdimen4\font\relax}
\providecommand{\BIBforeignlanguage}[2]{{%
\expandafter\ifx\csname l@#1\endcsname\relax
\typeout{** WARNING: IEEEtran.bst: No hyphenation pattern has been}%
\typeout{** loaded for the language `#1'. Using the pattern for}%
\typeout{** the default language instead.}%
\else
\language=\csname l@#1\endcsname
\fi
#2}}
\providecommand{\BIBdecl}{\relax}
\BIBdecl

\bibitem{MietznerCST09}
J.~Mietzner, R.~Schober, L.~Lampe, W.~H. Gerstacker, and P.~A. Hoeher,
  ``Multiple-antenna techniques for wireless communications -{A} comprehensive
  literature survey,'' \emph{IEEE Commun. Surveys $\&$ Tutorials}, vol.~11,
  no.~2, pp. 87--105, 2nd Quarter 2009.

\bibitem{Muirhead16}
D.~Muirhead, M.~A. Imran, and K.~Arshad, ``A survey of the challenges,
  opportunities and use of multiple antennas in current and future {5G} small
  cell base stations,'' \emph{IEEE Access}, vol.~4, pp. 2952--2964, 2016.

\bibitem{Saetal14}
A.~Sabharwal, P.~Schniter, D.~Guo, D.~W. Bliss, S.~Rangarajan, and R.~Wichman,
  ``In-band full-duplex wireless: Challenges and opportunities,'' \emph{IEEE J.
  Select. Areas Commun.}, vol.~32, no.~9, pp. 1637--1652, Feb. 2014.

\bibitem{ZhangCM15}
Z.~Zhang, X.~Chai, K.~Long, A.~V. Vasilakos, and L.~Hanzo, ``Full duplex
  techniques for {5G} networks: {S}elf-interference cancellation, protocol
  design, and relay selection,'' \emph{IEEE Commun. Magazine}, vol.~53, no.~5,
  pp. 128--137, May 2015.

\bibitem{DUPLO}
\BIBentryALTinterwordspacing
``System scenarios and technical requirements for full-duplex concept,''
  \emph{DUPLO project, Deliverable D1.1}. [Online]. Available: \url{at
  http://www.fp7-duplo.eu/index.php/deliverables.}
\BIBentrySTDinterwordspacing

\bibitem{Duarte:TWC:12}
M.~Duarte, C.~Dick, and A.~Sabharwal, ``Experiment-driven characterization of
  full-duplex wireless systems,'' \emph{IEEE Trans. Wireless Commun.}, vol.~11,
  no.~12, pp. 4296--4307, Dec. 2012.

\bibitem{Everett:14:TWC}
E.~Everett, A.~Sahai, and A.~Sabharwal, ``Passive self-interference suppression
  for full-duplex infrastructure nodes,'' \emph{IEEE Trans. Wireless Commun.},
  vol.~13, no.~2, pp. 680--694, Feb. 2014.

\bibitem{Riihonen-SP-11}
T.~Riihonen, S.~Werner, and R.~Wichman, ``Mitigation of loopback
  self-interference in full-duplex {MIMO} relays,'' \emph{IEEE Trans. Signal
  Process.}, vol.~59, no.~12, pp. 5983--5993, Dec. 2011.

\bibitem{Dan-SP-13}
D.~Nguyen, L.-N. Tran, P.~Pirinen, and M.~Latva-aho, ``Precoding for full
  duplex multiuser {MIMO} systems: Spectral and energy efficiency
  maximization,'' \emph{IEEE Trans. Signal Process.}, vol.~61, no.~16, pp.
  4038--4050, Aug. 2013.

\bibitem{Dan:TWC:14}
D.~Nguyen, L.-N. Tran, P.~Pirinen, and M.~Latva-aho, ``On the spectral
  efficiency of full-duplex small cell wireless systems,'' \emph{IEEE Trans.
  Wireless Commun.}, vol.~13, no.~9, pp. 4896--4910, Sept. 2014.

\bibitem{Tam:TCOM:16}
H.~H.~M. Tam, H.~D. Tuan, and D.~T. Ngo, ``Successive convex quadratic
  programming for quality-of-service management in full-duplex {MU-MIMO}
  multicell networks,'' \emph{IEEE Trans. Commun.}, vol.~64, no.~6, pp.
  2340--2353, June 2016.

\bibitem{LiTVT16}
Y.~Li, P.~Fan, L.~Anatolii, and L.~Liu, ``On the spectral and energy efficiency
  of full-duplex small cell wireless systems with massive {MIMO},'' \emph{IEEE
  Trans. Veh. Technol.}, to appear, available at
  http://ieeexplore.ieee.org/stamp/stamp.jsp?arnumber=7486140.

\bibitem{NguyenICC16}
D.~H.~N. Nguyen, L.~B. Le, and Z.~Han, ``Optimal uplink and downlink channel
  assignment in a full-duplex multiuser system,'' in \emph{Proc. IEEE Inter.
  Conf. Commun. (ICC 2016)}, May 2016, pp. 1--6.

\bibitem{SilvaTWC16}
J.~M.~B. da~Silva, G.~Fodor, and C.~Fischione, ``Spectral efficient and fair
  user pairing for full-duplex communication in cellular networks,'' \emph{IEEE
  Trans. Wireless Commun.}, vol.~15, no.~11, pp. 7578--7593, Nov. 2016.

\bibitem{AhnTWC16}
M.~Ahn, H.~B. Kong, H.~M. Shin, and I.~Lee, ``A low complexity user selection
  algorithm for full-duplex {MU-MISO} systems,'' \emph{IEEE Trans. Wireless
  Commun.}, vol.~15, no.~11, pp. 7899--7907, Nov. 2016.

\bibitem{RazaviyaynTSP14}
M.~Razaviyayn, M.~Baligh, A.~Callard, and Z.~Q. Luo, ``Joint user grouping and
  transceiver design in a {MIMO} interfering broadcast channel,'' \emph{IEEE
  Trans. Signal Process.}, vol.~62, no.~1, pp. 85--94, Jan. 2014.

\bibitem{SanjabiTSP14}
M.~Sanjabi, M.~Razaviyayn, and Z.~Q. Luo, ``Optimal joint base station
  assignment and beamforming for heterogeneous networks,'' \emph{IEEE Trans.
  Signal Process.}, vol.~62, no.~8, pp. 1950--1961, Apr. 2014.

\bibitem{Tse:book:05}
D.~Tse and P.~Viswanath, \emph{Fundamentals of Wireless Communication.}\hskip
  1em plus 0.5em minus 0.4em\relax Cambridge Univ. Press, UK, 2005.

\bibitem{Aardal:02}
K.~Aardal, R.~Weismantel, and L.~A. Wolsey, ``Non-standard approaches to
  integer programming,'' \emph{in Discrete Applied Mathematics}, pp. 5--74,
  2002.

\bibitem{WES06}
A.~Wiesel, Y.~Eldar, and S.~Shamai, ``Linear precoding via conic optimization
  for fixed {MIMO} receivers,'' \emph{IEEE Trans. Signal Process.}, vol.~54,
  no.~1, pp. 161--176, Jan. 2006.

\bibitem{Stephen}
S.~Boyd and L.~Vandenberghe, \emph{Convex Optimization}.\hskip 1em plus 0.5em
  minus 0.4em\relax Cambridge Univ. Press, UK, 2007.

\bibitem{Marks:78}
B.~R. Marks and G.~P. Wright, ``A general inner approximation algorithm for
  nonconvex mathematical programs,'' \emph{Operations Research}, vol.~26,
  no.~4, pp. 681--683, July-Aug. 1978.

\bibitem{Tuy-B-00}
H.~Tuy, \emph{Convex Analysis and Global Optimization}.\hskip 1em plus 0.5em
  minus 0.4em\relax Kluwer Academic, 2001.

\bibitem{Toh}
K.~C. Toh, M.~J. Todd, and R.~H. Tutuncu, ``{SDPT3-A} {M}atlab software package
  for semidefinite programming, version 1.3,'' \emph{Optimization Methods and
  Softw.}, vol.~11, pp. 545--581, Jan 1999.

\bibitem{MOSEK}
\BIBentryALTinterwordspacing
``I.~{MOSEK} aps,'' 2014. [Online]. Available: \url{at http://www.mosek.com.}
\BIBentrySTDinterwordspacing

\bibitem{Ben:2001}
A.~Ben-Tal and A.~Nemirovski, \emph{Lectures on {M}odern {C}onvex
  {O}ptimization.}\hskip 1em plus 0.5em minus 0.4em\relax Philadelphia:
  MPS-SIAM Series on Optimization, SIAM, 2001.

\bibitem{3GPP}
\emph{3GPP Technical Specification Group Radio Access Network, Evolved
  Universal Terrestrial Radio Access (E-UTRA): Further Advancements for E-UTRA
  Physical Layer Aspects (Release 9)}, document 3GPP TS 36.814 V9.0.0, 2010.

\bibitem{Bharadia13}
D.~Bharadia, E.~McMilin, and S.~Katti, ``Full duplex radios,'' in \emph{Proc.
  ACM SIGCOMM}, 2013.

\bibitem{Lofberg}
J.~Lofberg, ``{YALMIP}: {A} toolbox for modeling and optimization in
  {MATLAB},'' in \emph{Proc. IEEE Inter. Conf. Robotics and Auto. (IEEE Cat.
  No.04CH37508)}, Sept. 2004, pp. 284--289.

\end{thebibliography}

\end{document}